\documentclass[a4paper,12pt]{article}
\usepackage{amssymb}

\usepackage{contract}

\input{tcilatex}

\begin{document}

\setcounter{page}{0} \topmargin0pt \oddsidemargin5mm \renewcommand{%
\thefootnote}{\fnsymbol{footnote}} \newpage \setcounter{page}{0} 
\begin{titlepage}
\begin{flushright}
Berlin Sfb288 Preprint  \\
hep-th/0107015
\end{flushright}
\vspace{0.2cm}
\begin{center}
{\Large {\bf Form factors from free fermionic Fock fields,} }

{\Large {\bf the Federbush model} }

\vspace{0.8cm}
{\large  O.A.~Castro-Alvaredo  and  A.~Fring }

\vspace{0.2cm}
{Institut f\"ur Theoretische Physik, 
Freie Universit\"at Berlin,\\
Arnimallee 14, D-14195 Berlin, Germany }
\end{center}
\vspace{0.5cm}
 
\renewcommand{\thefootnote}{\arabic{footnote}}
\setcounter{footnote}{0}

\begin{abstract}
By representing the field content as well as the particle creation operators
in terms of  fermionic Fock operators, we compute the corresponding matrix elements 
of the Federbush model. Only when these matrix elements satisfy the form factor 
consistency equations involving anyonic factors of local commutativity, the corresponding
operators are local. We carry out the ultraviolet limit, analyse the
momentum space cluster properties and demonstrate how the Federbush model can
be obtained from the  $SU(3)_3$-homogeneous sine-Gordon model. We propose a new 
Lagrangian which on one hand constitutes a generalization of the Federbush model in
a Lie algebraic fashion and on the other a certain limit of the homogeneous sine-Gordon models.

\end{abstract}
\vfill{ \hspace*{-9mm}
\begin{tabular}{l}
\rule{6 cm}{0.05 mm}\\
Olalla@physik.fu-berlin.de\\
Fring@physik.fu-berlin.de
\end{tabular}}
\end{titlepage}
\newpage

\section{Introduction}

The analysis of the structure and properties, as well as the evaluation of
exact form factors, is one of the central problems in 1+1 dimensional
quantum field theories. One of the main reasons for their distinct role is
that they serve to compute very efficiently correlations functions of local
operators $\mathcal{O}(x)$. Instead of a perturbative expansion in the
coupling constant one may expand the correlation functions in terms of exact
expressions of n-particle form factors, that is the matrix element of a
local operator $\mathcal{O}(x)$ located at the origin between a
multiparticle in-state and the vacuum 
\begin{equation}
F_{n}^{\mathcal{O}|\mu _{1}\ldots \mu _{n}}(\theta _{1},\ldots ,\theta
_{n})\equiv \left\langle \mathcal{O}(0)\mathcal{\,}Z_{\mu _{1}}^{\dagger
}(\theta _{1})\ldots Z_{\mu _{n}}^{\dagger }(\theta _{n})\right\rangle _{%
\text{in}}\,.  \label{1}
\end{equation}
The operators $Z_{\mu }^{\dagger }(\theta )$ are creation operators for a
particle of type $\mu $ as a function of the rapidity $\theta $.

Since the original proposal of this method to evaluate correlation functions 
\cite{Kar}, various schemes have been suggested to compute these objects.
One of the original approaches is modeled in spirit closely on the set up
for the determination of exact scattering matrices. It consists of solving a
system of consistency equations which have to hold for the n-particle form
factors based on some natural physical assumptions, like unitarity, crossing
and bootstrap fusing properties \cite{Kar,Smir} 
\begin{eqnarray}
F_{n}^{\mathcal{O}|\ldots \mu _{i}\mu _{i+1}\ldots }(\ldots ,\theta
_{i},\theta _{i+1},\ldots ) &=&F_{n}^{\mathcal{O}|\ldots \mu _{i+1}\mu
_{i}\ldots }(\ldots ,\theta _{i+1},\theta _{i},\ldots )S_{\mu _{i}\mu
_{i+1}}(\theta _{i,i+1})\,,  \label{W1} \\
F_{n}^{\mathcal{O}|\mu _{1}\ldots \mu _{n}}(\theta _{1}+2\pi i,\ldots
,\theta _{n}) &=&\gamma _{\mu _{1}}^{\mathcal{O}}\,F_{n}^{\mathcal{O}|\mu
_{2}\ldots \mu _{n}\mu _{1}}(\theta _{2},\ldots ,\theta _{n},\theta
_{1})\,\,,  \label{W2}
\end{eqnarray}
\begin{equation}
\!\!\!\limfunc{Res}_{{\small \bar{\theta}}\rightarrow {\small \theta }_{0}}%
{\small F}_{n+2}^{\mathcal{O}|\bar{\mu}\mu \mu _{1}\ldots \mu _{n}}{\small (%
\bar{\theta}+i\pi ,\theta }_{0}{\small ,\theta }_{1}{\small \ldots \theta }%
_{n}{\small )}=i(1-\gamma _{\mu }^{\mathcal{O}}\prod_{l=1}^{n}S_{\mu \mu
_{l}}(\theta _{0l})){\small F}_{n}^{\mathcal{O}|\mu _{1}\ldots \mu _{n}}%
{\small (\theta }_{1}{\small \ldots \theta }_{n}{\small ).}  \label{kin}
\end{equation}
Here $\gamma _{\mu }^{\mathcal{O}}$ is the factor of so-called local
commutativity defined through the equal time exchange relation of the local
operator $\mathcal{O}(x)$ and the field $\mathcal{O}_{\mu }(y)$ associated
to the particle creation operators $Z_{\mu }^{\dagger }(\theta )$%
\begin{equation}
\mathcal{O}_{\mu }(x)\mathcal{O}(y)=\gamma _{\mu }^{\mathcal{O}}\,\mathcal{O}%
(y)\,\mathcal{O}_{\mu }(x)\qquad \,\text{for\thinspace \thinspace\ \ \ }%
x^{1}>y^{1}\,.  \label{local}
\end{equation}
The factor $\gamma _{\mu }^{\mathcal{O}}$ is very often omitted in the
analysis or simply taken to be one, but it can be seen that already in the
Ising model it is needed to set up the equations consistently \cite{YZam}. A
consequence of its presence is that a frequently made statement has to be
revised, namely, that (\ref{W1})-(\ref{kin}) constitute operator independent
equations, which require as the only input the S-matrices $S_{ij}(\theta
_{ij})$ between particles of type $i$ and $j$ as a function of the rapidity
difference $\theta _{ij}\equiv \theta _{i}-\theta _{j}$. In the following
manuscript we demonstrate that apart from $\pm 1$, which already occur in
the literature, this factor can be a non-trivial phase. Thus the form factor
consistency equations contain also explicitly non-trivial properties of the
operators.

Trying to find solutions to these equations has been pursuit successfully
for many models and has led to the determination of closed exact expressions
for n-particle form factors for a wide class of local operators $\mathcal{O}%
(x)$, e.g. \cite{Kar,Smir}.

Alternatively some authors develop methods which borrow ideas which have
proven to be very powerful in the context of conformal field theory, where
the use of symmetries and their related algebras has led to a successful
determination of correlation functions \cite{Stan}. Yet, the most direct way
to compute the matrix elements in (\ref{1}) is to find explicit
representations for the operators $Z_{\mu }^{\dagger }(\theta )$ and $%
\mathcal{O}(x)$. For instance in the context of lattice models this is a
rather familiar situation and one knows how to compute matrix elements of
the type (\ref{1}) directly. The problem is then reduced to a purely
computational task (albeit non-trivial), which may, for instance, be solved
by well-known techniques of algebraic Bethe ansatz type, e.g. \cite{Bethe}.
In the context of field theory a similar way of attack to the problem has
been followed by exploiting a free field representation for the operators $%
Z_{\mu }^{\dagger }(\theta )$ and $\mathcal{O}(x)$, in form of Heisenberg
algebras or their q-deformed version. So far a successful computation of the
n-particle form factors with this approach is limited to a rather restricted
set of models and in particular for the sine-Gordon model, which is a model
extensively studied by means of other approaches \cite{Smir,BFKZ}, only the
free Fermion point can be treated successfully \cite{MJ,Pak} so far. One of
the main purpose of this manuscript is to advocate yet another approach,
namely the evaluation of the matrix elements (\ref{1}) based on an expansion
of the operators in the conventional fermionic Fock space. Recalling the
well-known fact that in 1+1 space-time dimensions the notions of spin and
statistics are not intrinsic, it is clear that both approaches are
legitimate. Since the model we mainly consider in this manuscript, the
Federbush model, is closely related to complex free Fermions the usage of
fermionic Fock operators seems natural. Nonetheless, we expect this
procedure to hold in more generality and to allow an extension to other
models.

Our manuscript is organized as follows: In section 2 we recall an explicit
fermionic free field representation for the particle creation operators $%
Z_{\mu }^{\dagger }(\theta )$ occurring in (\ref{1}) valid for all diagonal
scattering matrices. In section 3 we treat the complex free Fermion, we
provide a generic expression for a potentially local operator and specialize
it to particular operators whose form factors we directly compute, namely
the order and disorder field and various components of the energy-momentum
tensor. In section 4 we extend this analysis to the Federbush model and show
in particular how it is related to homogeneous sine-Gordon (HSG) models on
the level of the scattering matrix. In addition we analyze the momentum
space cluster property. We pay special attention to the factor of local
commutativity. In section 5 we propose a Lie algebraic generalization of the
Federbush model. In section 6 we sustain the relation between the Federbush
and the HSG-models by carrying out the ultraviolet limit. We state our
conclusions in section 7.

\section{Fock space representation for the FZ-operators}

In order to proceed in the way as outlined above, we have to provide
explicit representations for the creation operators $Z_{\mu }^{\dagger
}(\theta )$ and the fields $\mathcal{O}(x)$. The former operators are
characterized by their braiding behaviour, i.e. when they are exchanged they
pick up the scattering matrix as a structure constant. We restrict our
considerations in this manuscript to theories in which backscattering is
absent, such that the exchange algebra for the $Z$-operators reads \cite{FZ1}
\begin{equation}
Z_{i}^{\dagger }(\theta _{i})Z_{j}^{\dagger }(\theta _{j})=S_{ij}(\theta
_{ij})Z_{j}^{\dagger }(\theta _{j})Z_{i}^{\dagger }(\theta _{i})=\exp [2\pi
i\delta _{ij}(\theta _{ij})]Z_{j}^{\dagger }(\theta _{j})Z_{i}^{\dagger
}(\theta _{i})\,.  \label{ZF}
\end{equation}
As indicated in (\ref{ZF}), the scattering matrix $S_{ij}(\theta _{ij})$ can
be expressed as a phase. Identical relations hold for the annihilation
operators, i.e. $Z^{\dagger }(\theta )\rightarrow Z(\theta )$ in (\ref{ZF}).
When we braid a creation and an annihilation operator the presence of an
additional central term was suggested in \cite{FZ2} 
\begin{equation}
Z_{i}(\theta _{i})Z_{j}^{\dagger }(\theta _{j})=S_{ij}(\theta
_{ij})Z_{j}^{\dagger }(\theta _{j})Z_{i}(\theta _{i})+2\pi \delta
_{ij}\delta (\theta _{i}-\theta _{j})\,,  \label{ZF2}
\end{equation}
which ensures that one recovers the usual (fermionic) bosonic
(anti)-commutation relations in the case ($S=-1$) $S=1$. The relations (\ref
{ZF})-(\ref{ZF2}) are commonly referred to as Faddeev-Zamolodchikov (FZ)
algebra. A representation for these operators in the bosonic Fock space was
first provided in \cite{AF} 
\begin{equation}
Z_{i}^{\dagger }(\theta )=\exp \left[ -i\int_{\theta }^{\infty }d\theta
^{\prime }\,\delta _{il}(\theta -\theta ^{\prime })a_{l}^{\dagger }(\theta
^{\prime })a_{l}(\theta ^{\prime })\right] a_{i}^{\dagger }(\theta )\,.
\label{real}
\end{equation}
By replacing a constant phase with the rapidity dependent phase $\delta
_{ij}(\theta )$ and turning the expression into a convolution with an
additional sum over $l$, the expression (\ref{real}) constitutes a
generalization of formulae found in the late seventies \cite{spin}, which
interpolate between bosonic and fermionic Fock spaces for arbitrary spin.
The latter construction may be viewed as a continuous version of a
Jordan-Wigner transformation \cite{JW}, albeit on the lattice the
commutation relations are not purely bosonic or fermionic, since certain
operators anti-commute at the same site but commute on different sites.
Alternatively, one may also replace the bosonic $a$'s in (\ref{real}) by
operators satisfying the usual fermionic anti-commutation relations 
\begin{equation}
\left\{ a_{i}(\theta ),a_{j}(\theta ^{\prime })\right\} =0\qquad \text{%
and\qquad }\left\{ a_{i}(\theta ),a_{j}^{\dagger }(\theta ^{\prime
})\right\} =2\pi \delta _{ij}\delta (\theta -\theta ^{\prime })\,
\label{ferm}
\end{equation}
and note that the relations (\ref{ZF}) are still satisfied \cite{Notas}. In
the following we want to work with this fermionic representation of the
FZ-algebra (\ref{ZF}). Having obtained a fairly simple realization for the $%
Z $-operators, we may now seek to represent the operator content of the
theory in the same space. In general, this is not known and we have to
resort to a study of explicit models at this stage.

\section{Complex free Fermions}

To illustrate the procedure, to fix some of our notations and to set the
scene for the Federbush model, we will commence with the free Fermion. Let
us consider $N$ complex (Dirac) free Fermions described as usual by the
Lagrangian density\footnote{%
We use the following conventions throughout the paper: 
\begin{eqnarray*}
x^{\mu } &=&(x^{0},x^{1}),\qquad p^{\mu }=(m\cosh \theta ,m\sinh \theta ) \\
g^{00} &=&-g^{11}=\varepsilon ^{01}=-\varepsilon ^{10}=1, \\
\gamma ^{0} &=&\left( 
\begin{array}{cc}
0 & 1 \\ 
1 & 0
\end{array}
\right) ,\quad \quad \gamma ^{1}=\left( 
\begin{array}{cc}
0 & 1 \\ 
-1 & 0
\end{array}
\right) ,\quad \gamma ^{5}=\gamma ^{0}\gamma ^{1}, \\
\psi _{\alpha } &=&\left( 
\begin{array}{c}
\psi _{\alpha }^{(1)} \\ 
\psi _{\alpha }^{(2)}
\end{array}
\right) ,\quad \bar{\psi}_{\alpha }=\psi _{\alpha }^{\dagger }\gamma ^{0}\,.
\end{eqnarray*}
} 
\begin{equation}
\mathcal{L}_{\text{FF}}=\sum_{\alpha =1}^{N}\bar{\psi}_{\alpha }(i\gamma
^{\mu }\partial _{\mu }-m_{\alpha })\psi _{\alpha }\,.  \label{fff}
\end{equation}
The associated equations of motion, i.e. the Dirac equations $(i\gamma ^{\mu
}\partial _{\mu }-m_{\alpha })\psi _{\alpha }=0$, may then of course be
solved with the help of the well-known Fourier decomposition of the complex
free Fermi fields 
\begin{equation}
\psi _{\alpha }(x)=\int \frac{dp_{\alpha }^{1}}{\sqrt{4\pi }p_{\alpha }^{0}}%
\left( a_{\alpha }(p)u_{\alpha }(p)e^{-ip_{\alpha }\cdot x}+a_{\bar{\alpha}%
}^{\dagger }(p)v_{\alpha }(p)e^{ip_{\alpha }\cdot x}\right) \,\qquad \alpha
=1,\ldots ,N.\,  \label{free}
\end{equation}
We abbreviated as usual $\sqrt{m_{\alpha }^{2}+p_{\alpha }^{2}}=p_{\alpha
}^{0}$ and employed the Weyl spinors 
\begin{equation}
u_{\alpha }(p)=\sqrt{\frac{m_{\alpha }}{2}}\left( 
\begin{array}{c}
e^{-\theta /2} \\ 
e^{\theta /2}
\end{array}
\right) \quad \quad \text{and\qquad }v_{\alpha }(p)=i\sqrt{\frac{m_{\alpha }%
}{2}}\left( 
\begin{array}{c}
e^{-\theta /2} \\ 
-e^{\theta /2}
\end{array}
\right) \,\,.  \label{WS}
\end{equation}

\noindent The amplitudes of the scattering matrices are simply $S_{\alpha
\alpha ^{\prime }}=-1$ for all combinations of $\alpha ,\alpha ^{\prime }$
and the FZ-algebra coincides by construction with the Clifford algebra (\ref
{ferm}), that is $Z_{\alpha }(\theta )=a_{\alpha }(\theta )$.

A further property, which we want to exploit here and in the next section,
is the $U(1)$-symmetry of the Lagrangian $\mathcal{L}_{\text{FF}}$, that is
changing 
\begin{equation}
\psi _{\alpha }(x)\rightarrow \eta _{\alpha }\psi _{\alpha }(x),
\label{symc}
\end{equation}
with $\eta _{\alpha }\in U(1)$ leaves the Lagrangian in (\ref{fff})
invariant. This simple symmetry will allow an a priori judgement about
vanishing form factors.

\subsection{Form factors of some local operators}

Let us now define a prototype auxiliary field 
\begin{eqnarray}
\chi _{\kappa }^{\alpha }(x) &=&\frac{1}{4\pi ^{2}}\int d\theta d\theta
^{\prime }\left[ \kappa ^{\alpha }(\theta ,\theta ^{\prime })\left(
a_{\alpha }^{\dagger }(\theta )a_{\bar{\alpha}}^{\dagger }(\theta ^{\prime
})e^{i(p+p^{\prime })\cdot x}+a_{\alpha }(\theta )a_{\bar{\alpha}}(\theta
^{\prime })e^{-i(p+p^{\prime })\cdot x}\right) \right.  \nonumber \\
&&\left. +\kappa ^{\alpha }(\theta ,\theta ^{\prime }-i\pi )\left( a_{\bar{%
\alpha}}^{\dagger }(\theta )a_{\bar{\alpha}}(\theta ^{\prime
})e^{i(p-p^{\prime })\cdot x}-a_{\alpha }(\theta )a_{\alpha }^{\dagger
}(\theta ^{\prime })e^{-i(p-p^{\prime })\cdot x}\right) \right] \,.
\label{aux}
\end{eqnarray}
This field is essentially bilinear in the free Fermi fields up to the
function $\kappa (\theta ,\theta ^{\prime })$, whose precise expression,
which gives the field its individual characteristic, we will leave generic
for the time being. The properties of this function, like $\kappa (\theta
-i\pi ,\theta ^{\prime }-i\pi )=\kappa (\theta ,\theta ^{\prime })$, as well
as the form of the space-time dependent exponentials, are dictated by the
crossing in (\ref{1}). It means bringing consistently some of the particle
creation operators $Z_{\mu }^{\dagger }(\theta )$ to the left of the
operator $\mathcal{O}(0)$ introduces these constraints. Fields of this
nature appear already in \cite{SMJ}. We now want to compute the matrix
element of a general operator composed out of $\chi _{\kappa }^{\alpha }(x)$ 
\begin{equation}
\mathcal{O}^{\chi _{\kappa }^{\alpha }}(x)=\mathbf{:}e^{\chi _{\kappa
}^{\alpha }(x)}\mathbf{:\,.}  \label{aux1}
\end{equation}
The direct computation of matrix elements related to these fields is
straightforward by employing Wick's first theorem\footnote{%
The difference between the product of some linear operators and its normal
ordered product has to be a c-number determined by all possible
contractions, i.e. for the linear operators $A,B,C,\ldots $ holds $ABC\ldots
-\mathbf{:}ABC\ldots \mathbf{:}=$ sum over all possible contractions (see
e.g. \cite{IZ}).}. Noting that the contribution from the normal ordered part
is of course zero, since all annihilation and creation operators are brought
to the left and right, respectively, we obtain for instance \ 
\begin{eqnarray}
\tilde{F}_{2}^{\chi _{\kappa }^{\alpha }|\bar{\alpha}\alpha }(\theta
_{1},\theta _{2}) &=&\frac{1}{4\pi ^{2}}\int d\theta d\theta ^{\prime
}\kappa ^{\alpha }(\theta ,\theta ^{\prime })\left( a_{\alpha }(\theta )a_{%
\bar{\alpha}}(\theta ^{\prime })a_{\bar{\alpha}}^{\dagger }(\theta
_{1})a_{\alpha }^{\dagger }(\theta _{2})\tallcon{12}{3}{11}\medcon{12}{5}{8}%
\right)  \nonumber \\
&=&\int d\theta d\theta ^{\prime }\kappa ^{\alpha }(\theta ,\theta ^{\prime
})\delta (\theta -\theta _{2})\delta (\theta ^{\prime }-\theta _{1})=\kappa
^{\alpha }(\theta _{2},\theta _{1})\,.
\end{eqnarray}
Proceeding in this way to higher particle numbers, we compute 
\begin{equation}
\tilde{F}_{2n}^{\mathcal{O}^{\chi _{\kappa }^{\alpha }}|n\times \bar{\alpha}%
\alpha }(\theta _{1},\ldots ,\theta _{2n})=\frac{1}{n!}\int d\theta
_{1}^{^{\prime }}\ldots d\theta _{2n}^{^{\prime }}\prod_{i=1}^{n}\kappa
^{\alpha }(\theta _{2i-1}^{\prime },\theta _{2i}^{\prime })\det \mathcal{D}%
^{2n}\,\,,  \label{even}
\end{equation}
where $\mathcal{D}^{\ell }$ is a rank $\ell $ matrix whose entries are given
by 
\begin{equation}
\mathcal{D}_{ij}^{\ell }=\cos ^{2}[(i-j)\pi /2]\delta (\theta _{i}^{^{\prime
}}-\theta _{j})\,,\qquad 1\leq i,j\leq \ell \,.
\end{equation}
We used the identity 
\begin{eqnarray}
&&a_{\alpha }(\theta _{n}^{\prime })\cdots a_{\bar{\alpha}}(\theta
_{3}^{\prime })a_{\alpha }(\theta _{2}^{\prime })a_{\bar{\alpha}}(\theta
_{1}^{\prime })a_{\bar{\alpha}}^{\dagger }(\theta _{1})a_{\alpha }^{\dagger
}(\theta _{2})a_{\bar{\alpha}}^{\dagger }(\theta _{3})\cdots a_{\alpha
}^{\dagger }(\theta _{n})\medcon{16}{2}{5}\tallcon{18}{1}{10}%
\xtallcon{22}{2}{17}\xxtallcon{25}{1}{24}+  \nonumber \\
&&a_{\alpha }(\theta _{n}^{\prime })\cdots a_{\bar{\alpha}}(\theta
_{3}^{\prime })a_{\alpha }(\theta _{2}^{\prime })a_{\bar{\alpha}}(\theta
_{1}^{\prime })a_{\bar{\alpha}}^{\dagger }(\theta _{1})a_{\alpha }^{\dagger
}(\theta _{2})a_{\bar{\alpha}}^{\dagger }(\theta _{3})\cdots a_{\alpha
}^{\dagger }(\theta _{n})\medcon{16}{2}{11}\tallcon{18}{1}{10}%
\xtallcon{22}{2}{11}\xxtallcon{25}{1}{24}+\cdots =\text{Det}\,\mathcal{D}%
^{n}.\,\,\quad \quad
\end{eqnarray}
A further generic field, which we want to study and which, in contrast to $%
\mathcal{O}^{\chi _{\kappa }^{\alpha }}$, now possesses non-vanishing matrix
elements with an odd particle number is 
\begin{equation}
\hat{\mathcal{O}}^{\chi _{\kappa }^{\alpha }}(x)=\mathbf{:}\hat{\psi}%
_{\alpha }(x)e^{\chi _{\kappa }^{\alpha }(x)}\mathbf{:\,.}  \label{aux2}
\end{equation}
This field involves the fermionic field with the spinor structure stripped
off 
\begin{equation}
\hat{\psi}_{\alpha }(x)=\int \frac{dp_{\alpha }^{1}}{2\pi p_{\alpha }^{0}}%
\left( a_{\alpha }(p)e^{-ip_{\alpha }\cdot x}+a_{\bar{\alpha}}^{\dagger
}(p)e^{ip_{\alpha }\cdot x}\right) \,.\,
\end{equation}
Similarly as before we compute the matrix elements 
\begin{equation}
\tilde{F}_{2n}^{\hat{\mathcal{O}}^{\chi _{\kappa }^{\alpha }}|\alpha
(n\times \bar{\alpha}\alpha )}(\theta _{1},\ldots ,\theta _{2n+1})=\frac{1}{%
n!}\int d\theta _{1}^{^{\prime }}\ldots d\theta _{2n+1}^{^{\prime
}}\prod_{i=1}^{n}\kappa ^{\alpha }(\theta _{2i}^{\prime },\theta
_{2i+1}^{\prime })\det \mathcal{D}^{2n+1}\,.  \label{odd}
\end{equation}
Note that $\mathcal{O}^{\chi _{\kappa }^{\alpha }}(x)$ and $\hat{\mathcal{O}}%
^{\chi _{\kappa }^{\alpha }}(x)$ are in general non-local operators, in the
sense that it is not guaranteed that they (anti)-commute for space-like
separations, i.e. $[\mathcal{O}(x),\mathcal{O}^{\prime }(y)]=0$ for $%
(x-y)^{2}<0$. At the same time $\tilde{F}_{n}^{\mathcal{O}}$ is just the
matrix element as defined on the r.h.s. of (\ref{1}) and not yet a form
factor of a local field, in the sense that it satisfies the consistency
equations (\ref{W1})-(\ref{kin}), which imply locality of $\mathcal{O}$%
\footnote{%
A rigorous proof of this statement to hold in generality is still an open
issue.}. In order to distinguish between this two different situations we
denote matrix elements in general by $\tilde{F}_{n}^{\mathcal{O}}$ and form
factors of local operators by\ $F_{n}^{\mathcal{O}}$. For instance, as a
consequence of the monodromy equation (\ref{W2}), a necessary condition for
these two functions to coincide for $\chi _{\kappa }^{\alpha }(x)$ is 
\begin{equation}
\kappa ^{\alpha }(\theta ,\theta ^{\prime }+2\pi i)=-\gamma _{\bar{\alpha}%
}^{\chi _{\kappa }^{\alpha }}\,\kappa ^{\alpha }(\theta ,\theta ^{\prime
})\,\,.
\end{equation}
Before specifying the functions $\kappa $ more concretely such that the
corresponding $\mathcal{O}$'s become local, we would like to compare briefly
the generic operators of the type (\ref{aux}), (\ref{aux1}) and (\ref{aux2})
with some general expressions for ``local'' operators which appear in the
literature \cite{Lask,Notas,Bert}. We carry out this argument in generality
without restriction to a concrete model. Let us restore in equation (\ref{1}%
) the space-time dependence, multiply the equation from the left with the
bra-vector $\left\langle Z_{\mu _{n}}^{\dagger }(\theta _{n})\ldots Z_{\mu
_{1}}^{\dagger }(\theta _{1})\right| $ and introduce the necessary amount of
sums and integrals over the complete states such that one can identify the
identity operator $\Bbb{I}$%
\begin{eqnarray*}
&&\!\!\sum\Sb n=1\ldots \infty  \\ \mu _{1}\ldots \mu _{n}  \endSb %
\int\limits_{-\infty }^{\infty }\frac{d\theta _{1}\ldots d\theta _{n}}{%
n!(2\pi )^{n}}F_{n}^{\mathcal{O}|\mu _{1}\ldots \mu _{n}}(\theta _{1}\ldots
\theta _{n})\left\langle Z_{\mu _{n}}^{\dagger }(\theta _{n})\ldots Z_{\mu
_{1}}^{\dagger }(\theta _{1})\right| e^{-i\sum_{j}p_{j}\cdot x} \\
&=&\!\!\sum\Sb n=1\ldots \infty  \\ \mu _{1}\ldots \mu _{n}  \endSb %
\int\limits_{-\infty }^{\infty }\frac{d\theta _{1}\ldots d\theta _{n}}{%
n!(2\pi )^{n}}\left\langle \mathcal{O}(x)\mathcal{\,}Z_{\mu _{1}}^{\dagger
}(\theta _{1})\ldots Z_{\mu _{n}}^{\dagger }(\theta _{n})\right\rangle
\left\langle Z_{\mu _{n}}^{\dagger }(\theta _{n})\ldots Z_{\mu
_{1}}^{\dagger }(\theta _{1})\right| \\
&=&\left\langle \mathcal{O}(x)\right. \mathcal{\,}\Bbb{I\,}.
\end{eqnarray*}
Cancelling the vacuum in the first and last line, and noting that we can
replace the product of operators, which is left over also by its normal
ordered version, we obtain the expression defined originally in \cite{Lask} 
\begin{equation}
\tilde{\!\mathcal{O}}(x)=\!\!\!\!\sum\Sb n=1\ldots \infty  \\ \mu _{1}\ldots
\mu _{n}  \endSb \int\limits_{-\infty }^{\infty }\frac{d\theta _{1}\ldots
d\theta _{n}}{n!(2\pi )^{n}}F_{n}^{\mathcal{O}|\mu _{1}\ldots \mu
_{n}}(\theta _{1}\ldots \theta _{n}):Z_{\mu _{n}}^{\dagger }(\theta
_{n})\ldots Z_{\mu _{1}}^{\dagger }(\theta
_{1}):e^{-i\sum\limits_{j}p_{j}\cdot x}  \label{crap}
\end{equation}
Hence this field is simply an inversion of (\ref{1}). From its very
construction it is clear that $\tilde{\mathcal{O}}(x)$ is a meaningful field
in the weak sense, that is acting on an in-state we will recover by
construction the form factor related to $\mathcal{O}(x)$. In addition, one
may also construct the well-known expression of the two-point correlation
function expanded in terms of form factors, as stated in \cite{Lask}.
However, it is also clear that $\tilde{\mathcal{O}}(x)\neq \mathcal{O}(x),$
simply by comparing (\ref{crap}) and the explicit expressions for some local
fields occurring in the free fermionic theory, e.g. (\ref{aux}), (\ref{aux1}%
) and (\ref{aux2}). The reason is that acting on an in-state with the latter
expressions the form factors are generated in a non-trivial Wick contraction
procedure, whereas when doing the same with (\ref{crap}) the Wick
contractions will be trivial. Therefore general statements and conclusions
drawn from an analysis made on $\tilde{\mathcal{O}}(x)$ should be taken with
care. It is also needless to say that from a practical point of view the
expression (\ref{crap}) is rather empty, since the expressions of the form
factors $F_{n}^{\mathcal{O}|\mu _{1}\ldots \mu _{n}}(\theta _{1}\ldots
\theta _{n})$ themselves are usually not known and their determination is in
general a quite non-trivial task. In \cite{Lask,Notas,Bert} the integration
in the formula (\ref{crap}) is a rather artificial contour integration which
takes care about analytic continuations of values of $i\pi $. This does not
seem to be a fundamental feature, since it remains completely obscure how to
incorporate bound states in this manner.

Let us now return to our concrete analysis by specifying $\kappa $.

\subsubsection{The order and disorder field}

Having in mind to proceed to the Federbush model, we will restrict ourselves
from now on to the case of two complex Fermions, i.e. $N=2$ in the
Lagrangian (\ref{fff}). The free fermionic theory possesses some very
distinct fields, namely the disorder and order fields 
\begin{equation}
\mu _{\alpha }(x)=\mathbf{:}e^{\omega _{\alpha }(x)}\mathbf{:\qquad }\text{%
and \qquad }\sigma _{\alpha }(x)=\mathbf{:}\hat{\psi}_{\alpha }(x)\mu
_{\alpha }(x)\mathbf{:},\quad \alpha =1,2,  \label{ggg}
\end{equation}
respectively. The names for these fields result from the ultraviolet limit,
see also section 5, since then they flow to their equivalent counterparts in
the conformal field theory \cite{BPZ}, namely to primary fields with scaling
dimension $1/16$. We introduced here the fields 
\begin{equation}
\omega _{\alpha }(x)=\chi _{\kappa }^{\alpha }(x),\quad \text{with\qquad }%
\kappa ^{1}(\theta ,\theta ^{\prime })=-\kappa ^{2}(-\theta ,-\theta
^{\prime })=\frac{i}{2}\frac{e^{-\frac{1}{2}(\theta -\theta ^{\prime })}}{%
\cosh \frac{1}{2}(\theta -\theta ^{\prime })}\quad \,.  \label{k}
\end{equation}
Admittedly, the precise form of the fields $\omega _{\alpha }(x)$ appears to
be slightly unmotivated at this stage. However, we will provide a better
rational for this in the next section, where we see that they originate by
relating a so-called triple normal ordering procedure for a field, which can
be constructed directly from the Fourier decomposition of the free Fermi
fields (\ref{free}), to another one associated with the usual Wick normal
ordering. It will turn out that the field $\omega _{\alpha }(x)$ emerges as
the limit of a Federbush model field to the free fermionic theory, i.e. $%
\lim_{\lambda \rightarrow 1/2}\Omega _{\alpha }^{\lambda }(x)=\omega
_{\alpha }(x)$, see equation (\ref{om}).

Let us now compute the form factors related to the above mentioned fields $%
\mu _{\alpha }(x)$ and $\sigma _{\alpha }(x)$. Using the particular form of $%
\kappa ^{\alpha }(\theta ,\theta ^{\prime })$ as defined in (\ref{k}), we
compute the integrals in (\ref{even}) and obtain a closed expression for the
n-particle form factors of the disorder operators 
\begin{eqnarray}
F_{2n}^{\mu _{1}|n\times \bar{1}1}(\theta _{1},\ldots ,\theta _{2n})
&=&(-1)^{n}F_{2n}^{\mu _{2}|n\times \bar{2}2}(-\theta _{1},\ldots ,-\theta
_{2n})  \nonumber \\
F_{2n}^{\mu _{\bar{1}}|n\times \bar{1}1}(-\theta _{1},\ldots ,-\theta _{2n})
&=&(-1)^{n}F_{2n}^{\mu _{\bar{2}}|n\times \bar{2}2}(\theta _{1},\ldots
,\theta _{2n})  \nonumber \\
&=&i^{n}2^{n-1}\sigma _{n}(\bar{x}_{1},\bar{x}_{3},\ldots ,\bar{x}_{2n-1})%
\mathcal{B}_{n,n}\,,  \label{555}
\end{eqnarray}
with 
\begin{equation}
\mathcal{B}_{n,m}=\frac{\prod\limits_{1\leq i<j\leq n}(\bar{x}_{2i-1}^{2}-%
\bar{x}_{2j-1}^{2})\prod\limits_{1\leq i<j\leq m}(x_{2i}^{2}-x_{2j}^{2})}{%
\prod\limits_{1\leq i<j\leq n+m}(u_{i}+u_{j})}=\frac{\det \mathcal{V}%
^{m}(x^{2})\det \mathcal{V}^{n}(\bar{x}^{2})}{\det \mathcal{W}^{n+m}(u)}\,.
\label{habibi}
\end{equation}
Associated with the particles and anti-particles we introduced here the
quantities $x_{i}=\exp (\theta _{i})$ and $\bar{x}_{i}=\exp (\theta _{i})$,
respectively. The variable $u_{i}$ can be either of them. We also employed
the elementary symmetric polynomials $\sigma _{k}(x_{1},\ldots ,x_{n})$,
defined as 
\begin{equation}
\sigma _{k}(x_{1},\ldots ,x_{n})=\sum_{l_{1}<\ldots <l_{k}}x_{l_{1}}\ldots
x_{l_{k}}\,\,\,\,\,
\end{equation}
(see e.g. \cite{Don} for more properties), the Vandermonde determinant of
the rank $\ell $ matrix $\mathcal{V}^{\ell }$ whose entries are given by 
\begin{equation}
\mathcal{V}_{ij}^{\ell }(x)=(x_{j})^{i-1},\qquad 1\leq i,j\leq \ell \,\,
\end{equation}
and the determinant of the rank $\ell -1$ matrix $\mathcal{W}^{\ell -1}$
with entries 
\begin{equation}
\mathcal{W}_{ij}^{\ell -1}(x)=\sigma _{2i-j}(x_{1},\ldots ,x_{\ell
})\,,\qquad 1\leq i,j\leq \ell \,-1\,.
\end{equation}
The relations between $F_{2n}^{\mu _{1}|n\times \bar{1}1}$ and $F_{2n}^{\mu
_{2}|n\times \bar{2}2}$ as stated in (\ref{555}) follow most transparently
from (\ref{k}) and (\ref{even}). One may easily verify that the expression (%
\ref{555}) indeed satisfies the consistency equations (\ref{W1})-(\ref{kin})
with $\gamma _{\bar{\alpha}}^{\mu _{\alpha }}=-1$ for $\alpha =1,2$. We
justify this choice in the next section by carrying out the Federbush model $%
\rightarrow $ two complex free Fermion limit. Noting that $\mu _{\alpha }(x)$
is invariant with respect to the symmetry property (\ref{symc}), it follows
immediately that 
\begin{equation}
F_{k+l}^{\mu _{\alpha ^{\prime }}|\alpha \alpha \ldots \alpha \alpha \beta
\beta \ldots \beta \beta }(\theta _{1},\ldots \theta _{k},\theta
_{k+1}\ldots ,\theta _{k+l})=0,\,\,\,\,\,\text{ }\alpha \neq \bar{\beta}%
\text{,\thinspace \thinspace }k\neq l\text{,\thinspace }\,\alpha ^{\prime
}=1,2,\bar{1},\bar{2}\,\,\,.
\end{equation}
This means that, up to a re-ordering of the particles, the expressions
reported in (\ref{555}) are in fact the only non-vanishing form factors
related to $\mu _{\alpha }(x)$ for $\alpha \in \{1,2,\bar{1},\bar{2}\}$.

In a similar way we compute the n-particle form factors of the order
operator 
\begin{eqnarray}
F_{2n+1}^{\sigma _{1}|1(n\times \bar{1}1)}(\theta _{1},\ldots ,\theta
_{2n+1}) &=&(-1)^{n}F_{2n+1}^{\sigma _{2}|2(n\times \bar{2}2)}(-\theta
_{1},\ldots ,-\theta _{2n+1})  \nonumber \\
F_{2n+1}^{\sigma _{\bar{1}}|1(n\times \bar{1}1)}(-\theta _{1},\ldots
,-\theta _{2n+1}) &=&(-1)^{n}F_{2n+1}^{\sigma _{\bar{2}}|2(n\times \bar{2}%
2)}(\theta _{1},\ldots ,\theta _{2n+1})  \nonumber \\
&=&i^{n}2^{n-1}\sigma _{n}(\bar{x}_{1},\ldots ,\bar{x}_{2n-1})\mathcal{B}%
_{n,n+1},  \label{6}
\end{eqnarray}
As a consistency check, one may once again verify that (\ref{6}) fulfills
the form factor equations (\ref{W1})-(\ref{kin}) with $\gamma _{\bar{\alpha}%
}^{\sigma _{\alpha }}=1$ for $\alpha =1,2$. Again, we postpone the
justification of this choice to the next section by carrying out the
Federbush model $\rightarrow $ two complex free Fermion limit. Noting that $%
\sigma _{\alpha }(x)$ $\rightarrow \eta _{\alpha }\sigma _{\alpha }(x)$ by (%
\ref{symc}), it follows immediately that 
\begin{equation}
F_{k+l}^{\sigma _{\alpha }|\alpha \alpha \ldots \alpha \alpha \beta \beta
\ldots \beta \beta }(\theta _{1},\ldots ,\theta _{k},\theta _{k+1}\ldots
,\theta _{k+l})=0\qquad \text{for }\alpha \neq \bar{\beta}\text{,\thinspace
\thinspace }k\neq l+1\,.
\end{equation}

Of course, this way of proceeding also works for the real free Fermion and
one may recover the well-known expressions of the literature \cite
{SMJ,BKW,YZam}. As a difference to our previous computations, however, we
have to take care of more contributions in the contraction procedure.
Keeping the form of $\kappa ^{\alpha }(\theta ,\theta ^{\prime })$ as
defined in (\ref{k}), but taking $\alpha =\bar{\alpha}$, we compute for
instance \ 
\begin{eqnarray}
F_{2}^{\mu }(\theta _{1},\theta _{2}) &=&\frac{1}{4\pi ^{2}}\int d\theta
d\theta ^{\prime }\kappa (\theta ,\theta ^{\prime })\left( a(\theta
)a(\theta ^{\prime })a^{\dagger }(\theta _{1})a^{\dagger }(\theta _{2})%
\tallcon{12}{4}{11}\medcon{12}{6}{9}+a(\theta )a(\theta ^{\prime
})a^{\dagger }(\theta _{1})a^{\dagger }(\theta _{2})\tallcon{12}{4}{9}%
\medcon{12}{6}{11}\right)  \nonumber \\
&=&\int d\theta d\theta ^{\prime }\kappa (\theta ,\theta ^{\prime })[\delta
(\theta -\theta _{2})\delta (\theta ^{\prime }-\theta _{1})-\delta (\theta
-\theta _{1})\delta (\theta ^{\prime }-\theta _{2})]  \nonumber \\
&=&i\tanh \frac{\theta _{12}}{2}\,.
\end{eqnarray}
Proceeding in this way to the higher n-particle form factors we only have to
replace in (\ref{even}) the matrix $\mathcal{D}^{\ell }$ with $\tilde{%
\mathcal{D}}^{\ell }$, whose entries are $\tilde{\mathcal{D}}_{ij}^{\ell
}=\delta (\theta _{i}^{^{\prime }}-\theta _{j})$. Computing the integrals we
get 
\begin{equation}
F_{2n}^{\mu }(\theta _{1},\ldots ,\theta _{2n})=i^{n}\func{Pf}(\mathcal{A}%
)=i^{n}\sqrt{\det \mathcal{A}}=i^{n}\prod_{1\leq i,j\leq 2n}\tanh \frac{%
\theta _{ij}}{2}\,,  \label{55}
\end{equation}
where $\mathcal{A}$ is an anti-symmetric ($2n\times 2n$)-matrix whose
entries are given by $\mathcal{A}_{ij}=\tanh \theta _{ij}/2$ and $\func{Pf}$
denotes its Pfaffian\footnote{%
Denoting the permutation group of $2n$ indices by $S_{2n}$ and the signature
of the permutation $\pi $ by $sgn(\pi )$, the Pfaffian of a matrix $\mathcal{%
A}$ is defined as 
\[
\func{Pf}(\mathcal{A})=\frac{1}{2^{n}n!}\sum\limits_{\pi \in S_{2n}}sgn(\pi
)\prod_{i=1}^{n}\mathcal{A}_{\pi (2i-1),\pi (2i)}\,. 
\]
}. In a similar way we compute the n-particle form factors of the order
operator 
\begin{equation}
F_{2n+1}^{\sigma }(\theta _{1},\ldots ,\theta _{2n+1})=i^{n}\func{Pf}(%
\mathcal{A})=i^{n}\prod_{1\leq i,j\leq 2n+1}\tanh \frac{\theta _{ij}}{2}\,.
\label{66}
\end{equation}
Expressions of the type (\ref{55}) and (\ref{66}) can be found already in
the first paper of \cite{SMJ}. The product expressions for $F_{2n}^{\mu }$
and $F_{2n+1}^{\sigma }$ were also derived in \cite{BKW} and \cite{YZam},
respectively, by means of solving the form factor consistency equations (\ref
{W1})-(\ref{kin}).

\subsubsection{The energy momentum-tensor}

A further field which plays an important role in any theory is the
energy-momentum tensor, which for the free Fermion in our normalization
simply reads 
\begin{equation}
T_{\;\;\nu }^{\mu }=2i(\mathbf{:}\bar{\psi}_{1}\gamma ^{\mu }\partial _{\nu
}\psi _{1}:\mathbf{+\,\mathbf{:}}\bar{\psi}_{2}\gamma ^{\mu }\partial _{\nu
}\psi _{2}\mathbf{\mathbf{:)}.}
\end{equation}
With the help of equation (\ref{free}) we compute easily 
\begin{equation}
T_{\;\;}^{\mu \nu }=\chi _{t_{1}^{\mu \nu }}+\chi _{t_{2}^{\mu \nu }}\,,
\end{equation}
using 
\begin{equation}
t_{\alpha }^{0\mu }(\theta ,\tilde{\theta})=-2\pi im_{\alpha }(p_{\alpha
})^{\mu }\sinh \frac{\theta +\tilde{\theta}}{2},\,\,\text{ \quad }t_{\alpha
}^{1\mu }(\theta ,\tilde{\theta})=-2\pi im_{\alpha }(p_{\alpha })^{\mu
}\cosh \frac{\theta +\tilde{\theta}}{2}\,,
\end{equation}
where we recall the definition of $\chi _{\kappa }(x)$ from (\ref{aux}). A
specially distinct role is played by the trace of the energy-momentum
tensor, since on one hand it is directly proportional to the operator which
breaks the conformal invariance \cite{Cardy} and on the other hand it occurs
explicitly in various computations associated to the ultraviolet limit like
the c-theorem \cite{ZamC} and the $\Delta $-sum rule \cite{DSC} (see section
5). It acquires the explicit form 
\begin{equation}
T_{\;\;\mu }^{\mu }=2im_{1}\mathbf{:}\bar{\psi}_{1}\psi _{1}\mathbf{:}%
+2im_{2}\mathbf{:}\bar{\psi}_{2}\psi _{2}\mathbf{:=}\chi _{t_{1}}+\chi
_{t_{2}}\,,\quad \quad t_{\alpha }(\theta ,\tilde{\theta})=2\pi im_{\alpha
}^{2}\sinh \frac{\tilde{\theta}-\theta }{2}\,.
\end{equation}
It is clear that only the two-particle form factor can be different from
zero and we compute it in an analogous way as in the previous section, that
is using Fourier decomposition (\ref{free}) with subsequent contractions, 
\begin{eqnarray}
F_{2}^{T^{0\mu }|\bar{\alpha}\alpha }(\theta ,\tilde{\theta}) &=&-2\pi
im_{\alpha }p^{\mu }\sinh \frac{\theta +\tilde{\theta}}{2},\quad \\
F_{2}^{T_{\;}^{1\mu }|\bar{\alpha}\alpha }(\theta ,\tilde{\theta}) &=&-2\pi
im_{\alpha }p^{\mu }\cosh \frac{\theta +\tilde{\theta}}{2}, \\
F_{2}^{T_{\;\;\mu }^{\mu }|\bar{\alpha}\alpha }(\theta ,\tilde{\theta})
&=&F_{2}^{T_{\;\;\mu }^{\mu }|\alpha \bar{\alpha}}(\theta ,\tilde{\theta}%
)=-2\pi im_{\alpha }^{2}\sinh \frac{\theta -\tilde{\theta}}{2}\,.
\end{eqnarray}
When taking $\alpha =\bar{\alpha}$, these expressions coincide with the ones
which may be found in the literature for the real Fermion. As usual, we may
verify that various equations which hold for the operators themselves also
hold for the associated form factors. For instance, the conservation of the
energy-momentum tensor 
\begin{equation}
\partial _{\mu }T^{\mu \nu }=i\left[ \hat{P}_{\mu },T^{\mu \nu }\right] =0
\label{cons}
\end{equation}
is reflected by the fact that 
\begin{equation}
\left( p^{0}+\tilde{p}^{0}\right) F_{2}^{T^{\mu 0}|\bar{\alpha}\alpha
}=-\left( p^{1}+\tilde{p}^{1}\right) F_{2}^{T^{\mu 1}|\bar{\alpha}\alpha }\,.
\end{equation}
Here we used the explicit form of the momentum operator 
\begin{equation}
\hat{P}_{\mu }=\int_{-\infty }^{\infty }dx^{1}T_{\;\;\mu }^{0}=\sum_{\alpha
=1}^{2}\int \frac{dp_{\alpha }^{1}}{2\pi p_{\alpha }^{0}}\,(p_{\alpha
})_{\mu }\,(a_{\alpha }^{\dagger }(p)a_{\alpha }(p)-a_{\bar{\alpha}}(p)a_{%
\bar{\alpha}}^{\dagger }(p))  \label{P}
\end{equation}
when changing in (\ref{cons}) derivatives to commutators by means of the
Heisenberg equation of motion. It is then easy to verify that $[\hat{P}_{\mu
},a_{\alpha }^{\dagger }(p)]=(p_{\alpha })_{\mu }a_{\alpha }^{\dagger }(p)$
and $[\hat{P}_{\mu },a_{\alpha }(p)]=-(p_{\alpha })_{\mu }a_{\alpha }(p)$,
such that we verify explicitly 
\begin{equation}
\partial _{\mu }\chi _{\kappa }^{\alpha }(x)=i[\hat{P}_{\mu },\chi _{\kappa
}^{\alpha }(x)]\,\,,  \label{Hchi}
\end{equation}
which is of course what we expect. Equation (\ref{Hchi}) is a further
support for the consistency of the generic definition of $\chi _{\kappa
}^{\alpha }(x)\,$\ in (\ref{aux}).

\section{The Federbush Model}

The Federbush model \cite{Feder} was proposed forty years ago as a prototype
for an exactly solvable quantum field theory which obeys the Wightman axioms 
\cite{Wight,Ruij,USP}. Formally it is closely related to the massive
Thirring model \cite{Thirring}. It contains two different massive particles $%
\Psi _{1}$ and $\Psi _{2}$. A special feature of this model is that the
related vector currents $J_{\alpha }^{\mu }=\bar{\Psi}_{\alpha }\gamma ^{\mu
}\Psi _{\alpha }$, $\alpha \in \{1,2\}$, whose analogues occur squared in
the massive Thirring model, enter the Lagrangian density of the Federbush
model in a parity breaking manner 
\begin{equation}
\mathcal{L}_{\text{F}}=\sum_{\alpha =1,2}\bar{\Psi}_{\alpha }(i\gamma ^{\mu
}\partial _{\mu }-m_{\alpha })\Psi _{\alpha }-2\pi \lambda \varepsilon _{\mu
\nu }J_{1}^{\mu }J_{2}^{\nu }\,  \label{LFeder}
\end{equation}
due to the presence of the Levi-Civita pseudotensor $\varepsilon $. It is
then easy to verify that the related equations of motion 
\begin{equation}
(i\gamma ^{\mu }\partial _{\mu }-m_{1})\Psi _{1}=2\pi \lambda \varepsilon
_{\mu \nu }J_{2}^{\nu }\gamma ^{\mu }\Psi _{1},\quad (i\gamma ^{\mu
}\partial _{\mu }-m_{2})\Psi _{2}=2\pi \lambda \varepsilon _{\nu \mu
}J_{1}^{\nu }\gamma ^{\mu }\Psi _{2},  \label{Fequm}
\end{equation}
can be solved by 
\begin{equation}
\Psi _{1}=\vdots \exp (2\sqrt{\pi }i\lambda \phi _{2})\vdots \psi
_{1}\,=\Phi _{2}^{\lambda }\psi _{1},\quad \Psi _{2}=\vdots \exp (-2\sqrt{%
\pi }i\lambda \phi _{1})\vdots \psi _{2}\,=\Phi _{1}^{\lambda }\psi _{2},
\label{sol}
\end{equation}
if in addition the free bosonic fields $\phi _{\alpha }$ constitute
potentials for axial vector currents composed out of the free Fermions $\psi
_{\alpha }$ 
\begin{equation}
\frac{1}{\sqrt{\pi }}\partial _{\mu }\phi _{\alpha }=\varepsilon _{\nu \mu
}J_{\alpha }^{\nu }=\bar{\psi}_{\alpha }\gamma _{\mu }\gamma ^{5}\psi
_{\alpha },\qquad \lambda \neq 0,\quad \alpha =1,2\,\,.\,  \label{vec}
\end{equation}
The triple normal ordering in equation (\ref{sol}) is defined as $\mathbf{%
\vdots }e^{\kappa \phi }\mathbf{\vdots }=e^{\kappa \phi }/\left\langle
e^{\kappa \phi }\right\rangle $ for $\kappa $ being some constant. This is
very advantageous in the calculation of commutation relations, since one can
simply deal with ordinary operator relations instead of having to handle
messy Wick contractions. We stress that in case the coupling constant $%
\lambda $ vanishes, that is when $\mathcal{L}_{\text{F}}$ reduces to $%
\mathcal{L}_{\text{FF}}$ and the relations (\ref{Fequm}) correspond to two
decoupled Dirac equations, the relation (\ref{vec}) does not hold.

In order to compute the factors of local commutativity $\gamma _{\mu }^{%
\mathcal{O}}$, as defined in (\ref{local}), we need various
(anti)-commutation relations. The fields $\psi _{\alpha }(x)$ are complex
free (Dirac) Fermions of masses $m_{\alpha }$ and the fields $\phi _{\alpha
}(x)$ are free Bosons, such that for $\alpha ,\beta =1,2$ we trivially have 
\begin{eqnarray}
\lbrack \phi _{\alpha }(x),\phi _{\beta }(y)] &=&[\Phi _{\alpha }(x),\Phi
_{\beta }(y)]=[\phi _{\alpha }(x),\Phi _{\beta }(y)]=\{\psi _{\alpha
}(x),\psi _{\beta }(y)\}=0\,,\,\,  \label{444} \\
\{\psi _{\alpha }(x),\psi _{\beta }^{\dagger }(y)\} &=&\delta _{\alpha \beta
}\delta (x^{1}-y^{1})\,\,.
\end{eqnarray}
The commutation relations involving mixed expressions of $\psi _{\alpha }$
and $\phi _{\beta }$ are less obvious and in fact it is crucial to note that
these fields are not mutually local, that is $[\psi _{\alpha }(x),\phi
_{\beta }(y)]\neq 0$ for space-like separations, i.e. $(x-y)^{2}<0$.
Concretely we have the following equal time exchange relations for $\alpha
,\beta =1,2$%
\begin{eqnarray}
\lbrack \psi _{\alpha }(x),\phi _{\beta }(y)] &=&\sqrt{\pi }\delta _{\alpha
\beta }\Theta (x^{1}-y^{1})\psi _{\alpha }(x)\,,  \label{3c} \\
\psi _{\alpha }(x)\Phi _{\beta }^{\lambda }(y) &=&\Phi _{\beta }^{\lambda
}(y)\psi _{\alpha }(x)\,e^{2\pi i(-1)^{\beta }\lambda \delta _{\alpha \beta
}\Theta (x^{1}-y^{1})}\,, \\
-\psi _{\alpha }(x)\Psi _{\beta }(y) &=&\Psi _{\beta }(y)\psi _{\alpha
}(x)\,e^{-2\pi i(-1)^{\beta }\lambda \delta _{|\alpha -\beta |,1}\Theta
(x^{1}-y^{1})}\,, \\
\Psi _{\alpha }(x)\Phi _{\beta }^{\lambda }(y) &=&\Phi _{\beta }^{\lambda
}(y)\Psi _{\alpha }(x)\,e^{2\pi i(-1)^{\beta }\lambda \delta _{\alpha \beta
}\Theta (x^{1}-y^{1})}\,, \\
-\Psi _{\alpha }(x)\Psi _{\beta }(y) &=&\Psi _{\beta }(y)\Psi _{\alpha
}(x)\,e^{-2\pi i\lambda (-1)^{\beta }\delta _{|\alpha -\beta |,1}}.
\label{antic}
\end{eqnarray}

\noindent We used here the Heavyside step function $\Theta (x)$, defined as
usual as $\Theta (x>0)=1$, $\Theta (x<0)=0$ and $\Theta (0)=1/2$. One may
convince oneself easily that (\ref{3c}) is compatible with (\ref{vec}) and
that the remaining equations are straightforward consequences of (\ref{444}%
)-(\ref{3c}). Apart from this choice, which agrees with the one in \cite{LS}%
, one can also find in some places of the literature, e.g. \cite{STW}, that
in (\ref{3c}) the $\Theta $-functions are replaced by $\varepsilon
(x)/2=\Theta (x)-\Theta (-x)$. This is of course also compatible with (\ref
{vec}). However, an immediate consequence of our choice is that the
Federbush fields $\Psi _{\alpha }(x)$ are only mutually local if they are of
the same type $\alpha $, whereas when taking the $\varepsilon $-function
instead, they are mutually local for all values of $\alpha \,$\ and $\beta $%
. The different choices will of course lead to different factors of local
commutativity $\gamma $ and will therefore alter the consistency equations (%
\ref{W1})-(\ref{kin}). Arguing on the properties of these equations we
provide more reasoning for our choice below.

A further important implication of the fact that $\psi _{\alpha }$ and $\phi
_{\beta }$ are not mutually local is that the fields $\Psi _{\alpha }$ are
in different Borchers classes\footnote{%
An equivalence class of complete, local field systems is referred to as a
Borchers class. Its crucial property is that it characterizes completely the
scattering matrix without having to resort to particular fields. (For more
details see e.g. \cite{Haag} p.104, however, this notion is of no further
relevance for our concrete computations.)} as the free Fermion. Thus, there
is a chance for the existence of a nontrivial scattering matrix, which was
indeed found in \cite{Wight,STW}. In fact, we will now demonstrate that this
S-matrix can be obtained as a limit of a more complex model, that is the
homogeneous sine-Gordon (HSG) model.

\subsection{Federbush Models from HSG-models}

Ever since the equivalence between the massive Thirring- and the sine-Gordon
model was demonstrated \cite{Cole}, there have been various identifications
between different types of models. In a similar spirit we also want to show
now how a fermionic model is obtainable from a bosonic one, albeit in
contrast to the above situation our fermionic scattering matrix will be
constructed solely out of the asymptotic phases of a given scattering matrix 
\begin{equation}
\lim_{\theta \rightarrow \pm \infty }S_{ij}(\theta )=e^{\Delta _{ij}^{\pm
}\,}.  \label{sstart}
\end{equation}
Starting from a consistent solution to the crossing and unitarity relations 
\begin{equation}
S_{ab}^{jk}(\theta )S_{ba}^{kj}(-\theta )=1\quad \quad \text{and\qquad }%
S_{ab}^{jk}(\theta )=S_{\bar{b}a}^{\bar{k}j}(i\pi -\theta )\,,  \label{cu}
\end{equation}
one clearly has the constraint $\Delta _{ij}^{+}=-\Delta _{ji}^{-}=\Delta _{%
\bar{\jmath}i}^{-}$. This means that 
\begin{equation}
\hat{S}_{ij}=e^{\Delta _{ij}^{+}+\Delta _{ij}^{-}\,}  \label{sne}
\end{equation}
will also be a valid solution to the unitarity-crossing relations for $S$.
Having no rapidity dependence there is no bound state bootstrap equation to
be concerned about, such that (\ref{sne}) already constitutes a consistent
scattering matrix. Concretely we will now show that when taking $%
S_{ij}(\theta )$ in (\ref{sstart}) to be the scattering matrix of the $%
SU(3)_{3}$-HSG model, as found in \cite{HSGS}, the resulting S-matrix $%
\hat{S}_{ij}$ in (\ref{sne}) will be the one of the Federbush model at a
particular value of the coupling constant. In fact it can be shown that this
prescription leads to a much wider range of scattering matrices which can be
directly associated to a Lagrangian of a Federbush model generalized in a
Lie algebraic manner, see section 5. Let us recall now the scattering matrix
of the $SU(3)_{3}$-HSG model 
\begin{equation}
S^{SU(3)_{3}}(\theta )=\left( 
\begin{array}{cccc}
(2)_{\theta } & -(1)_{\theta } & -(-2)_{\theta }e^{-i\pi \tau } & 
(-1)_{\theta }e^{i\pi \tau } \\ 
-(1)_{\theta } & (2)_{\theta } & (-1)_{\theta }e^{i\pi \tau } & 
-(-2)_{\theta }e^{-i\pi \tau } \\ 
-(-2)_{\theta }e^{i\pi \tau } & (-1)_{\theta }e^{-i\pi \tau } & (2)_{\theta }
& -(1)_{\theta } \\ 
(-1)_{\theta }e^{-i\pi \tau } & -(-2)_{\theta }e^{i\pi \tau } & -(1)_{\theta
} & (2)_{\theta }
\end{array}
\right) .  \label{62}
\end{equation}
We abbreviated $(x)_{\theta }=\sinh \frac{1}{2}(\theta +i\pi x/3)/\sinh 
\frac{1}{2}(\theta -i\pi x/3)$ and $\tau =\pm 1/3$. For the rows and columns
we adopt here the ordering $\{1,\bar{1},2,\bar{2}\}$. We also took the
resonance parameters $\sigma $ of the HSG-model to be zero, since they will
not play any role in our further considerations. Computing now the limit
according to the above prescription we obtain 
\begin{equation}
\lim_{\theta \rightarrow \infty }[S_{ij}^{SU(3)_{3}}(\theta
)S_{ij}^{SU(3)_{3}}(-\theta )]=S^{\text{FB}}=-\left( 
\begin{array}{cccc}
1 & 1 & e^{-2\pi i\lambda } & e^{2\pi i\lambda } \\ 
1 & 1 & e^{2\pi i\lambda } & e^{-2\pi i\lambda } \\ 
e^{2\pi i\lambda } & e^{-2\pi i\lambda } & 1 & 1 \\ 
e^{-2\pi i\lambda } & e^{2\pi i\lambda } & 1 & 1
\end{array}
\right) \,.  \label{SFeder}
\end{equation}
We found it convenient to relate the parameter $\tau $ to the $\lambda $ in
the Lagrangian density (\ref{LFeder}) as $\tau =1-\lambda $. Then $S^{\text{%
FB}}$ corresponds to the scattering matrix derived in \cite{Wight,STW},
apart from the overall minus sign, which is due to the fact that we adopt
the convention that the particles are ordered in opposite order in the in-
and out-states, i.e. we include the statistics factor into the S-matrix.
After having taken the limit (\ref{SFeder}), the crossing and unitarity
equations also hold when we relax the constraint for $\tau $ and allow it to
take completely generic values different from $1/3$. Thus, whenever the
coupling constant $\lambda $ becomes an even integer the theory decouples
into a system of two free complex Fermions. From a Lagrangian point of view
we expect this kind of behaviour of course for vanishing $\lambda $.

Having specified the scattering matrix of the model, we are in the position
to state directly from (\ref{real}) a representation for the FZ-algebra. The
explicit version of (\ref{real}) then reads 
\begin{eqnarray}
Z_{1}^{\dagger }(\theta ) &=&\exp \left( -i\lambda \int_{\theta }^{\infty
}d\theta ^{\prime }\mathbf{:}\rho _{2}(\theta ^{\prime })\mathbf{:}\right)
a_{1}^{\dagger }(\theta )  \label{Z1} \\
Z_{\bar{1}}^{\dagger }(\theta ) &=&\exp \left( i\lambda \int_{\theta
}^{\infty }d\theta ^{\prime }\mathbf{:}\rho _{2}(\theta ^{\prime })\mathbf{:}%
\right) a_{\bar{1}}^{\dagger }(\theta ) \\
Z_{2}^{\dagger }(\theta ) &=&\exp \left( i\lambda \int_{\theta }^{\infty
}d\theta ^{\prime }\mathbf{:}\rho _{1}(\theta ^{\prime })\mathbf{:}\right)
a_{2}^{\dagger }(\theta ) \\
Z_{\bar{2}}^{\dagger }(\theta ) &=&\exp \left( -i\lambda \int_{\theta
}^{\infty }d\theta ^{\prime }\mathbf{:}\rho _{1}(\theta ^{\prime })\mathbf{:}%
\right) a_{\bar{2}}^{\dagger }(\theta )  \label{Z4}
\end{eqnarray}
with 
\begin{equation}
\rho _{\alpha }(\theta )=a_{\alpha }^{\dagger }(\theta )a_{\alpha }(\theta
)-a_{\bar{\alpha}}^{\dagger }(\theta )a_{\bar{\alpha}}(\theta )\,.
\end{equation}
We will now specify more concretely various local operators of the Federbush
model for which we want to compute the form factors explicitly by using the
fermionic free field representation (\ref{Z1})-(\ref{Z4}).

\subsection{Form factors of some local operators}

We compute now explicitly the bosonic fields $\phi _{\alpha }(x)$ by solving
equation (\ref{vec}) and express them in terms of our general formula (\ref
{aux}) 
\begin{equation}
\phi _{\alpha }(x)=\sqrt{\pi }\int_{-\infty }^{x^{1}}dx^{1}\mathbf{:}\Psi
_{\alpha }^{\dagger }\Psi _{\alpha }\mathbf{:}=\chi _{\tilde{\kappa}%
}^{\alpha }(x)\,\,\,\,\,\,\text{with\qquad }\tilde{\kappa}^{\alpha }(\theta
,\theta ^{\prime })=\frac{\pi ^{\frac{3}{2}}}{2\cosh \frac{1}{2}(\theta
-\theta ^{\prime })}\,.  \label{BOSE}
\end{equation}
A field closely related to $\phi _{\alpha }(x)$, but whose origin is far
less direct, is 
\begin{equation}
\Omega _{\alpha }^{\lambda }(x)=\chi _{\hat{\kappa}}^{\alpha }(x)\qquad 
\text{with\qquad }\hat{\kappa}^{1}(\theta ,\theta ^{\prime })=-\hat{\kappa}%
^{2}(-\theta ,-\theta ^{\prime })=\frac{i\sin (\pi \lambda )e^{-\lambda
(\theta -\theta ^{\prime })}}{2\cosh \frac{1}{2}(\theta -\theta ^{\prime })}%
\,.  \label{om}
\end{equation}
It is this field which constitutes the analogue to the auxiliary field
already used in the previous section. In view of the periodicity of the
scattering matrix (\ref{LFeder}), we may restrict the range of $\lambda $ to 
$\lambda \in (0,2)/1$. The special role of $\lambda =1$ was treated in more
detail in \cite{ST}. Important for our purposes is the value $\lambda =1/2$
for which the operator $\Omega _{\alpha }^{\lambda }(x)$ reduces to $\omega
_{\alpha }(x)$ as defined in equation (\ref{aux}).

\subsubsection{The order and disorder field}

In close relation to the free fermionic theory one may also introduce the
analogue fields to the disorder and order fields in the Federbush model 
\begin{equation}
\Phi _{\alpha }^{\lambda }(x)=\mathbf{:}\exp [\Omega _{\alpha }^{\lambda
}(x)]\,\mathbf{:\qquad }\text{and \qquad }\Sigma _{\alpha }^{\lambda }(x)=%
\mathbf{:}\hat{\psi}_{\alpha }(x)\mathbf{\,}\,\Phi _{\alpha }^{\lambda }(x)%
\mathbf{:\,.}  \label{problemilla}
\end{equation}
In \cite{LS}, Lehmann and Stehr showed the remarkable fact that the operator 
$\Phi _{\alpha }^{\lambda }(x)$, which is composed out of free Bosons,
occurring in (\ref{sol}) can be viewed in two equivalent ways. On one hand
it can be defined through a so-called triple ordered product and on the
other hand by means of a conventional fermionic Wick ordered expression 
\begin{equation}
\Phi _{\alpha }^{\lambda }(x)=\vdots \exp [-2\sqrt{\pi }i\lambda \phi
_{\alpha }(x)]\vdots =:\exp [\Omega _{\alpha }^{\lambda }(x)]\,\mathbf{%
:\quad .}  \label{problemon}
\end{equation}
Having again in mind to compute the factors of local commutativity $\gamma
_{\mu }^{\mathcal{O}}$, as defined in (\ref{local}), we need various equal
time exchange relations. With the help of (\ref{3c})-(\ref{antic}) we
compute 
\begin{eqnarray}
-\psi _{\alpha }(x)\Sigma _{\beta }^{\lambda }(y) &=&\Sigma _{\beta
}^{\lambda }(y)\psi _{\alpha }(x)\,e^{2\pi i(-1)^{\beta }\lambda \delta
_{\alpha \beta }\Theta (x^{1}-y^{1})}\,, \\
\Phi _{\alpha }^{\lambda }(x)\Sigma _{\beta }^{\lambda }(y) &=&\Sigma
_{\beta }^{\lambda }(y)\Phi _{\alpha }^{\lambda }(x)\,e^{2\pi i(-1)^{\beta
}\lambda \delta _{\alpha \beta }\Theta (x^{1}-y^{1})}\,, \\
-\Psi _{\alpha }(x)\Sigma _{\beta }^{\lambda }(y) &=&\Sigma _{\beta
}^{\lambda }(y)\Psi _{\alpha }(x)\,e^{2\pi i\lambda (-1)^{\beta }(\delta
_{\alpha \beta }\Theta (x^{1}-y^{1})-\delta _{|\alpha -\beta |,1}\Theta
(y^{1}-x^{1}))}\,, \\
\Sigma _{\alpha }^{\lambda }(x)\Sigma _{\beta }^{\lambda }(y) &=&\Sigma
_{\beta }^{\lambda }(y)\Sigma _{\alpha }^{\lambda }(x)\,e^{2\pi i(-1)^{\beta
}\lambda \delta _{\alpha \beta }}\,\,.  \label{fucked}
\end{eqnarray}
Having obtained the relevant exchange relations we can read off the factors
of local commutativity for the operators under consideration 
\begin{equation}
\gamma _{\alpha }^{\Phi _{\beta }^{\lambda }}=-\gamma _{\alpha }^{\Sigma
_{\beta }^{\lambda }}=e^{2\pi i(-1)^{\beta }\lambda \delta _{\alpha \beta
}}\quad \text{and\quad }\gamma _{\bar{\alpha}}^{\Phi _{\beta }^{\lambda
}}=-\gamma _{\bar{\alpha}}^{\Sigma _{\beta }^{\lambda }}=e^{-2\pi
i(-1)^{\beta }\lambda \delta _{\alpha \beta }}\quad .\,  \label{carallo}
\end{equation}
Note in particular, that for $\lambda \rightarrow 1/2$ we recover, as we
expect, the values corresponding to the two complex free Fermions 
\begin{equation}
\lim_{\lambda \rightarrow 1/2}\gamma _{\alpha }^{\Phi _{\alpha }^{\lambda
}}=\gamma _{\alpha }^{\mu _{\alpha }}=-1\quad \quad \text{and\quad \quad }%
\lim_{\lambda \rightarrow 1/2}\gamma _{\alpha }^{\Sigma _{\alpha }^{\lambda
}}=\gamma _{\alpha }^{\sigma _{\alpha }}=1\,\,.
\end{equation}

Having assembled all the ingredients, let us now turn to the explicit
computation of the n-particle form factors related to the field $\Phi
_{\alpha }^{\lambda }(x)$. Since $\mathcal{L}_{\text{FF}}$ respects the same
symmetry as $\mathcal{L}_{\text{F}}$, namely (\ref{symc}), it is an
immediate consequence that the only non-vanishing form factors of $\Phi
_{\alpha }^{\lambda }(x)$ have to involve an equal number of particles and
anti-particles $\alpha $ and $\bar{\alpha}$. That means 
\begin{equation}
F_{k+l}^{\Phi _{\alpha ^{\prime }}^{\lambda }|\alpha \alpha \ldots \alpha
\alpha \beta \beta \ldots \beta \beta }(\theta _{1},\ldots \theta
_{k},\theta _{k+1}\ldots ,\theta _{k+l})=0\,\,,\text{ }\alpha \neq \bar{\beta%
}\text{,\thinspace \thinspace }k\neq l\text{,\thinspace }\,\alpha ^{\prime
}=1,2,\bar{1},\bar{2}\,.  \label{keinProblem}
\end{equation}
Turning now to the non-vanishing form factors, we compute by employing again
Wick's theorem 
\begin{eqnarray}
F_{2}^{\Phi _{1}^{\lambda }|\bar{1}1}(\theta _{1},\theta _{2}) &=&\frac{1}{%
4\pi ^{2}}\int d\theta d\theta ^{\prime }\kappa ^{\alpha }(\theta ,\theta
^{\prime })\left( a_{\alpha }(\theta )a_{\bar{\alpha}}(\theta ^{\prime })Z_{%
\bar{\alpha}}^{\dagger }(\theta _{1})Z_{\alpha }^{\dagger }(\theta _{2})%
\tallcon{12}{3}{11}\medcon{12}{5}{8}\right)  \nonumber \\
&=&\frac{i\sin (\pi \lambda )e^{\lambda \theta _{12}}}{2\cosh \frac{1}{2}%
\theta _{12}}=F_{2}^{\Phi _{2}^{-\lambda }|\bar{2}2}(\theta _{1},\theta
_{2})\,.  \label{phi}
\end{eqnarray}
Note, that in the contraction of an $a_{\alpha }(\theta )$- and a $Z_{\alpha
}^{\dagger }(\theta )$-operator there is no contribution from the
exponential term inside the $Z_{\alpha }^{\dagger }(\theta )$, since it
involves always particles of a different type than $\alpha $, see (\ref{Z1}%
)-(\ref{Z4}). Proceeding again in the same way as in the previous section,
we obtain as closed expressions for the n-particle form factors 
\begin{eqnarray}
&&F_{2n}^{\Phi _{1}^{\lambda }|n\times \bar{1}1}(\bar{x}_{1},x_{2}\ldots 
\bar{x}_{2n-1},x_{2n})=(-1)^{n}F_{2n}^{\Phi _{2}^{-\lambda }|n\times \bar{2}%
2}(\bar{x}_{1},x_{2}\ldots \bar{x}_{2n-1},x_{2n})=\qquad  \nonumber \\
&&F_{2n}^{\Phi _{\bar{1}}^{-\lambda }|n\times \bar{1}1}(\bar{x}%
_{1},x_{2}\ldots \bar{x}_{2n-1},x_{2n})=(-1)^{n}F_{2n}^{\Phi _{\bar{2}%
}^{\lambda }|n\times \bar{2}2}(\bar{x}_{1},x_{2}\ldots \bar{x}%
_{2n-1},x_{2n})=  \nonumber \\
&&\qquad \quad \qquad \qquad i^{n}2^{n-1}\sin ^{n}(\pi \lambda )\sigma _{n}(%
\bar{x}_{1}\ldots \bar{x}_{2n-1})^{\lambda +\frac{1}{2}}\sigma
_{n}(x_{2}\ldots x_{2n})^{\frac{1}{2}-\lambda }\mathcal{B}_{n,n}\,.\mathcal{%
\quad }  \label{chocho}
\end{eqnarray}
We may now convince ourselves, that the expressions for $F_{2n}^{\Phi
_{\alpha }^{\lambda }|n\times \bar{\alpha}\alpha }$ indeed satisfy the
consistency equations (\ref{W1})-(\ref{kin}). The first two equations are
rather obvious to check and we will not report this computation here, but
the verification of the kinematic residue equation (\ref{kin}) deserves
mentioning 
\begin{eqnarray}
&&\!\!\!\!\!\limfunc{Res}_{{\small x}\rightarrow {\small \bar{x}}%
}F_{2n+2}^{\Phi _{1}^{\lambda }|(n+1)\times \bar{1}1}(-\bar{x},x,\bar{x}%
_{1},x_{2},\ldots ,\bar{x}_{2n-1},x_{2n})=  \nonumber \\
&&\!\!\!\!\!2^{n}i^{n+1}\sin ^{n+1}(\pi \lambda )\sigma _{n+1}(-x,\bar{x}%
_{1}\ldots \bar{x}_{2n-1})^{\lambda +\frac{1}{2}}\sigma _{n+1}(x,x_{2}\ldots
x_{2n})^{\frac{1}{2}-\lambda }\limfunc{Res}_{{\small x}\rightarrow {\small 
\bar{x}}}\mathcal{B}_{n+1,n+1}=  \nonumber \\
&&\!\!\!\!\!i\left[ 1-\gamma _{1}^{\Phi _{1}^{\lambda
}}\prod_{k=1}^{2n}S_{1k}\right] F_{2n}^{\Phi _{1}^{\lambda }|n\times \bar{1}%
1}(\bar{x}_{1},\ldots \bar{x}_{2n-1},x_{2n})\,\,.\,\ \ \ \quad  \label{polla}
\end{eqnarray}
Recalling the definition of $\mathcal{B}_{n,n}$ of (\ref{habibi}), we used $%
\limfunc{Res}_{x\rightarrow \bar{x}}\mathcal{B}_{n+1,n+1}=-x^{-1}\mathcal{B}%
_{n,n}$ and the value for $\gamma _{1}^{\Phi _{1}^{\lambda }}$ from (\ref
{carallo}). Note the factor $\sin (\pi \lambda )$, which was originally
found in \cite{ST}, and which appears in our presentation in (\ref{om})
relatively unmotivated, is absolutely crucial for the validity of (\ref
{polla}).

Similarly we evaluate the matrix elements of $\Sigma _{\alpha }^{\lambda }$ 
\begin{eqnarray}
&&\tilde{F}_{2n+1}^{\Sigma _{1}^{\lambda }|1(n\times \bar{1}1)}(\theta
_{1},\ldots ,\theta _{2n+1})=(-1)^{n}\tilde{F}_{2n+1}^{\Sigma _{2}^{-\lambda
}|2(n\times \bar{2}2)}(\theta _{1},\ldots ,\theta _{2n+1})=  \nonumber \\
&&\tilde{F}_{2n+1}^{\Sigma _{\bar{1}}^{-\lambda }|1(n\times \bar{1}%
1)}(\theta _{1},\ldots ,\theta _{2n+1})=(-1)^{n}\tilde{F}_{2n+1}^{\Sigma _{%
\bar{2}}^{\lambda }|2(n\times \bar{2}2)}(\theta _{1},\ldots ,\theta _{2n+1})=%
\frac{(2i)^{n}}{2}\sin ^{n}(\pi \lambda )  \nonumber \\
&&\,\times \frac{\sigma _{n}(\bar{x}_{2}\ldots \bar{x}_{2n})^{\lambda +\frac{%
1}{2}}}{\sigma _{n}(x_{1}\ldots x_{2n+1})^{\lambda -\frac{1}{2}}}%
\prod\limits_{1\leq i<j\leq n}(\bar{x}_{2i}-\bar{x}_{2j})\sum_{k}\frac{%
i^{k+1}\prod\limits_{j<l;j,l\neq k}(x_{j}-x_{l})}{(x_{k})^{\frac{1}{2}%
-\lambda }\,\prod\limits_{j\neq k}\prod\limits_{l}(x_{j}+\bar{x}_{l})}%
\,.\,\,\,  \label{85}
\end{eqnarray}
However, the expressions of $\tilde{F}_{2n+1}^{\Sigma _{\alpha }^{\lambda
}|\alpha (n\times \bar{\alpha}\alpha )}$ only satisfy the consistency
equations (\ref{W1})-(\ref{kin}) for $\lambda =1/2$. This reflects the fact
that $\Sigma _{\alpha }^{\lambda }(x)$ is only a local operator for this
value of $\lambda $, see equation (\ref{fucked}). Thus, the equations (\ref
{W1})-(\ref{kin}) ``know'' about the locality properties of the operator
involved.

As we already commented above, part of the operator content of the Federbush
model reduces to the one of the complex fermionic theory. We may check
explicitly that the same limit is respected by the form factors 
\begin{eqnarray}
\lim_{\lambda \rightarrow 1/2}F_{2n}^{\Phi _{\alpha }^{\lambda }|n\times 
\bar{\alpha}\alpha }(\theta _{1},\ldots ,\theta _{2n}) &=&F_{2n}^{\mu
_{\alpha }|n\times \bar{\alpha}\alpha }(\theta _{1},\ldots ,\theta
_{2n}),\quad \quad \\
\lim_{\lambda \rightarrow 1/2}\tilde{F}_{2n+1}^{\Sigma _{\alpha }^{\lambda
}|\alpha (n\times \bar{\alpha}\alpha )}(\theta _{1},\ldots ,\theta _{2n+1})
&=&F_{2n+1}^{\sigma _{\alpha }|\alpha (n\times \bar{\alpha}\alpha )}(\theta
_{1},\ldots ,\theta _{2n+1})\,.
\end{eqnarray}
Note, however, that since the real fermionic theory can not be obtained
directly from the Federbush model, see also section 3.1.1., we also do not
recover, as we expect, the same expressions for the form factors when the
particles are taken to be self-conjugate.

\subsubsection{The Federbush fields}

Let us now compute the form factors of some fields which occur explicitly in
the Federbush model. From the expressions of the previous section the form
factors for the Federbush fields follow easily 
\begin{eqnarray}
F_{2n+1}^{\Psi _{1}|(n\times \bar{2}2)1}(\theta _{1},\ldots ,\theta _{2n+1})
&=&\sqrt{\pi }F_{2n}^{\Phi _{2}^{\lambda }|n\times \bar{2}2}(\theta
_{1},\ldots ,\theta _{2n})\,u_{1}(\theta _{2n+1})\,\,,  \label{c1} \\
F_{2n+1}^{\Psi _{2}|(n\times \bar{1}1)2}(\theta _{1},\ldots ,\theta _{2n+1})
&=&\sqrt{\pi }F_{2n}^{\Phi _{1}^{\lambda }|n\times \bar{1}1}(\theta
_{1},\ldots ,\theta _{2n})\,u_{2}(\theta _{2n+1})\,\,.
\end{eqnarray}
Recall the definition of the Weyl spinors $u_{\alpha }(\theta )\,$\ from
equation (\ref{WS}). It is clear that for each component these fields
satisfy the form factor consistency equations. As already mentioned in
section 3.1.2. and as is quite common in the literature, e.g. \cite
{Zamocorr,deter}, third reference in \cite{BFKZ} etc., one may verify that
various equations which hold for the operators are also satisfied by the
related form factors. However, one should be aware that such relations also
hold for the matrix elements $\tilde{F}$, which do not yet satisfy the
consistency equations (\ref{W1})-(\ref{kin}). Hence, the only conclusion one
may draw from such comparisons is a relative consistency amongst the
solutions obtained. Such arguments do not serve as a stringent
identification of the operators, albeit they give an indication. We
illustrate this statement with the following simple computation. Let us take
the fields as defined in (\ref{BOSE}) and evaluate directly by Wick
contracting 
\begin{eqnarray}
\tilde{F}_{2}^{\partial _{0}\phi _{\alpha }|\bar{\alpha}\alpha }(\theta ,%
\tilde{\theta}) &=&\sqrt{\pi }\tilde{F}_{2}^{J_{\alpha }^{1}|\bar{\alpha}%
\alpha }(\theta ,\tilde{\theta})=-i\pi ^{\frac{3}{2}}m_{\alpha }\cosh \frac{%
\theta +\tilde{\theta}}{2},\quad \alpha =1,2\,\,,  \label{b1} \\
\tilde{F}_{2}^{\partial _{1}\phi _{\alpha }|\bar{\alpha}\alpha }(\theta ,%
\tilde{\theta}) &=&-\sqrt{\pi }\tilde{F}_{2}^{J_{\alpha }^{0}|\bar{\alpha}%
\alpha }(\theta ,\tilde{\theta})=i\pi ^{\frac{3}{2}}m_{\alpha }\sinh \frac{%
\theta +\tilde{\theta}}{2},\quad \alpha =1,2\,\,.  \label{b2}
\end{eqnarray}
This confirms precisely the conservation equations (\ref{vec}) on the level
of the matrix elements. However, it is also easy to see that the expressions
(\ref{b1}) and (\ref{b2}) are not yet solutions of the form factor
consistency equations (\ref{W1})-(\ref{kin}). In principle, these equations
together with the Dirac equation already ensure that the $\Psi _{\alpha }$
are solutions of the equations of motion (\ref{Fequm}). Nonetheless, it is
instructive to verify (\ref{Fequm}) explicitly. Using still the
representation (\ref{BOSE}), we compute 
\begin{eqnarray}
\tilde{F}_{3}^{\varepsilon _{\mu \nu }J_{2}^{\nu }\gamma ^{\mu }\Psi _{1}|%
\bar{2}21}(\theta _{1},\theta _{2},\theta _{3})\!\!\! &=&\!\!\!i\pi ^{\frac{3%
}{2}}m_{2}u_{1}(\theta _{1}+\theta _{2}-\theta _{3})\,\,  \label{91} \\
\tilde{F}_{3}^{\gamma ^{\mu }\partial _{\mu }\Psi _{1}|\bar{2}21}(\theta
_{1},\theta _{2},\theta _{3})\!\!\! &=&\!\!\!\frac{i\pi ^{\frac{5}{2}%
}\lambda (m_{2}u_{1}(\theta _{13}+\theta _{1})+m_{2}u_{1}(\theta
_{13}+\theta _{1})+m_{1}u_{1}(\theta _{3}))}{\cosh \frac{1}{2}\theta _{12}}%
\quad \,\,\, \\
\tilde{F}_{3}^{\Psi _{1}|\bar{2}21}(\theta _{1},\theta _{2},\theta
_{3})\!\!\! &=&\!\!\!\frac{i\pi ^{\frac{5}{2}}\lambda }{\cosh \frac{1}{2}%
\theta _{12}}u_{1}(\theta _{3})\,\,.  \label{93}
\end{eqnarray}
Assembling these expressions, we confirm directly the validity of (\ref
{Fequm}) at the level of the three particle matrix elements. We expect of
course this property also to holds for higher orders. It is easy to check
that (\ref{91})-(\ref{93}) do not constitute solutions of the equations (\ref
{W1})-(\ref{kin}), in particular (\ref{c1}) does not reduce to (\ref{93}).
Thus on one hand we see that formal operator equations do not serve as a
conclusive means of operator identification and we therefore need
alternative arguments such as the ultraviolet limit in section 5 etc. On the
other hand this underlines further the need for the introduction of the
field $\Omega _{\alpha }^{\lambda }(x)$.

\subsubsection{The energy momentum tensor}

The energy-momentum tensor for the Federbush model has been computed in \cite
{SH}. Its evaluation involved a small subtlety, since the one obtained
directly from the Lagrangian does not lead to the correct Poincar\'{e}
generators, such as (\ref{P}). This could be fixed in the usual way by
exploiting the ambiguity in the definition. Essential for our purposes is
once again the trace, which is 
\begin{equation}
T_{\;\;\mu }^{\mu }=2im_{1}\mathbf{:}\bar{\Psi}_{1}\Psi _{1}\mathbf{:+}%
2im_{2}\mathbf{:}\bar{\Psi}_{2}\Psi _{2}\mathbf{:}\quad .
\end{equation}
Using the representation (\ref{sol}) for the Federbush fields, we compute
the only non-vanishing form factor for $T_{\;\;\mu }^{\mu }$ to \ 
\begin{equation}
F_{2}^{T_{\;\;\mu }^{\mu }|\bar{\alpha}\alpha }(\theta ,\tilde{\theta}%
)=F_{2}^{T_{\;\;\mu }^{\mu }|\alpha \bar{\alpha}}(\theta ,\tilde{\theta}%
)=-2\pi im_{\alpha }^{2}\sinh \frac{\theta -\tilde{\theta}}{2}\,\,.
\label{EMTF}
\end{equation}
This means the function is the same as the one for the complex free Fermion.

\subsection{Momentum space cluster properties}

As a consequence of Weinberg's power counting theorem one has also a further
property of form factors which involves the structure of the operators
themselves, namely the momentum space cluster property, see e.g. \cite{Kar}
some reasoning on this. It serves on one hand as a consistency check for
possible solutions of (\ref{W1})-(\ref{kin}) and on the other as a
construction principle for new solutions, e.g. \cite{CF1}. It states that
whenever some of the rapidities, say $\kappa $, are shifted to plus or minus
infinity, the $n$-particle form factor related to a local operator $\mathcal{%
O}$ factorizes into a $\kappa $ and an ($n-\kappa $)-particle form factor
which are possibly related to different types of operators $\mathcal{O}%
^{\prime }$ and $\mathcal{O}^{\prime \prime }$. Introducing the translation
operator $T_{a}^{\vartheta }$ which acts on a function of $n$ variables as 
\begin{equation}
T_{a}^{\vartheta }\,f(\theta _{1},\ldots ,\theta _{a},\ldots ,\theta
_{n})\,\mapsto \,f(\theta _{1},\ldots ,\theta _{a}+\vartheta ,\ldots ,\theta
_{n})\,\,\,
\end{equation}
and the operators 
\begin{equation}
\bar{\mathcal{T}}_{a,b}^{\pm }=\lim_{\vartheta \rightarrow \infty
}\prod_{p=a}^{b}T_{2p-1}^{\pm \vartheta },\qquad \qquad \text{and}\qquad
\qquad \mathcal{T}_{a,b}^{\pm }=\lim_{\vartheta \rightarrow \infty
}\prod_{p=a}^{b}T_{2p}^{\pm \vartheta },
\end{equation}
the statement of momentum space cluster decomposition reads 
\begin{equation}
\bar{\mathcal{T}}_{a,\kappa }^{\pm }\mathcal{T}_{a,\kappa }^{\pm }F_{n}^{%
\mathcal{O}}(\theta _{1}\ldots \theta _{n})\sim F_{2(\kappa -a+1)}^{\mathcal{%
O}^{\prime }}(\theta _{2a-1}\ldots \theta _{2\kappa })F_{n-2(\kappa -a+1)}^{%
\mathcal{O}^{\prime \prime }}(\theta _{1}\ldots \theta _{2a-2},\theta
_{2\kappa +1}\ldots \theta _{n})\,\,.  \label{cluster}
\end{equation}
Of course, we could have defined the product of $\bar{\mathcal{T}}_{a,\kappa
}^{\pm }\mathcal{T}_{a,\kappa }^{\pm }$ to be just one operator, but it will
be convenient for us to distinguish the shifts in even and odd positions of
the particles. Let us now see the effect of the action of these operators on
the various functions which build up our form factor solutions, see (\ref
{555}), (\ref{6}) and (\ref{chocho}). We compute

\begin{equation}
\bar{\mathcal{T}}_{1,\kappa }^{\pm }\mathcal{T}_{1,\zeta }^{\pm }\left[ 
\frac{\sigma _{n}(\bar{x}_{1}\ldots \bar{x}_{2n-1})^{\lambda +\frac{1}{2}}}{%
\sigma _{m}(x_{2}\ldots x_{2m})^{\lambda -\frac{1}{2}}}\right] \sim e^{\pm
\lambda \vartheta (\kappa -\zeta )\pm \vartheta \frac{(\kappa +\zeta )}{2}}%
\left[ \frac{\sigma _{n}(\bar{x}_{1}\ldots \bar{x}_{2n-1})^{\lambda +\frac{1%
}{2}}}{\sigma _{m}(x_{2}\ldots x_{2m})^{\lambda -\frac{1}{2}}}\right] ,
\label{11}
\end{equation}
and 
\begin{equation}
\bar{\mathcal{T}}_{1,\kappa }^{\pm }\mathcal{T}_{1,\zeta }^{\pm }\mathcal{B}%
_{n,m}\sim \mathcal{B}_{\kappa ,\varsigma }\mathcal{B}_{n-\kappa
,m-\varsigma }\left\{ \QTATOP{e^{\vartheta (\zeta -\kappa )\left[ \frac{%
\kappa -\zeta }{2}-n+m\right] -\vartheta \frac{(\kappa +\zeta )}{2}}\left[ 
\frac{\sigma _{\kappa }(\bar{x}_{1}\ldots \bar{x}_{2\kappa -1})}{\sigma
_{\varsigma }(x_{2}\ldots x_{2\varsigma })}\right] ^{n-m+\varsigma -\kappa }%
}{e^{-\vartheta \frac{(\kappa -\zeta )^{2}}{2}+\vartheta \frac{(\kappa
+\zeta )}{2}}\left[ \frac{\sigma _{n-\kappa }(\bar{x}_{2\kappa +1}\ldots 
\bar{x}_{2n-1})}{\sigma _{m-\varsigma }(x_{2\varsigma +2}\ldots x_{2m})}%
\right] ^{\kappa -\varsigma }}\right. \,\,.  \label{22}
\end{equation}
In order not to overload our symbols, we have slightly abused here the
notation. Whereas in (\ref{habibi}) the $x_{i},\bar{x}_{j}$-dependence of $%
\mathcal{B}_{n,m}$ always start at $i,j=1$, in (\ref{22}) the dependence of $%
\mathcal{B}_{n-\kappa ,m-\varsigma }$ for the minus shift is the same as in
the corresponding factor for the symmetric polynomials. Besides the explicit
functional dependence on the r.h.s. of (\ref{11}) \ and (\ref{22}) it is
instructive to consider at first the leading order behaviour 
\begin{equation}
\bar{\mathcal{T}}_{1,\kappa }^{\pm }\mathcal{T}_{1,\zeta }^{\pm }\left[ 
\frac{\sigma _{n}(\bar{x}_{1}\ldots \bar{x}_{2n-1})^{\lambda +\frac{1}{2}}}{%
\sigma _{m}(x_{2}\ldots x_{2m})^{\lambda -\frac{1}{2}}}\right] \mathcal{B}%
_{n,m}\sim \left\{ \QTATOP{e^{\vartheta (\zeta -\kappa )\left[ \frac{\kappa
-\zeta }{2}-n+m-\lambda \right] }}{e^{\vartheta (\zeta -\kappa )\left[ \frac{%
\kappa -\zeta }{2}+\lambda \right] }}\right. \,.
\end{equation}
From this we see directly that in general the final expression will tend to
zero, unless $\zeta =\kappa $, $|\zeta -\kappa |=2\lambda $ or $|\zeta
-\kappa \pm 2|=2\lambda $, by noting that our solutions only allow $%
|n-m|=0,1 $. So, let us now collect the functional dependences in equations (%
\ref{11}) and (\ref{22}) and see how our form factor solutions combine under
clustering to new form factors. We compute 
\begin{eqnarray}
\bar{\mathcal{T}}_{1,\kappa }^{\pm }\mathcal{T}_{1,\kappa }^{\pm
}F_{2n}^{\Phi _{\alpha }^{\lambda }|n\times \bar{\alpha}\alpha }\!\!\! &\sim
&\!\!\!F_{2\kappa }^{\Phi _{\alpha }^{\lambda }|\kappa \times \bar{\alpha}%
\alpha }(\theta _{1}\ldots \theta _{2\kappa })F_{2(n-\kappa )}^{\Phi
_{\alpha }^{\lambda }|(n-\kappa )\times \bar{\alpha}\alpha }(\theta
_{2\kappa +1}\ldots \theta _{2n})\,, \\
\bar{\mathcal{T}}_{1,\kappa }^{\pm }\mathcal{T}_{1,\kappa }^{\pm
}F_{2n}^{\Phi _{\bar{\alpha}}^{\lambda }|n\times \bar{\alpha}\alpha }\!\!\!
&\sim &\!\!\!F_{2\kappa }^{\Phi _{\bar{\alpha}}^{\lambda }|\kappa \times 
\bar{\alpha}\alpha }(\theta _{1}\ldots \theta _{2\kappa })F_{2(n-\kappa
)}^{\Phi _{\bar{\alpha}}^{\lambda }|(n-\kappa )\times \bar{\alpha}\alpha
}(\theta _{2\kappa +1}\ldots \theta _{2n})\,, \\
\bar{\mathcal{T}}_{1,\kappa +1}^{\pm }\mathcal{T}_{1,\kappa }^{\pm
}F_{2n}^{\Phi _{\alpha }^{\pm \frac{1}{2}}|n\times \bar{\alpha}\alpha
}\!\!\!\! &\sim &\!\!\!\!F_{2\kappa +1}^{\sigma _{\bar{\alpha}}|(\kappa
\times \bar{\alpha}\alpha )\bar{\alpha}}(\theta _{1}\ldots \theta _{2\kappa
+1})F_{2(n-\kappa )-1}^{\sigma _{\alpha }|[(n-\kappa )\times \bar{\alpha}%
\alpha ]\alpha }(\theta _{2\kappa +2}\ldots \theta _{2n}),\quad \quad \,\,\,
\\
\bar{\mathcal{T}}_{1,\kappa }^{\pm }\mathcal{T}_{1,\kappa +1}^{\pm
}F_{2n}^{\Phi _{\alpha }^{\mp \frac{1}{2}}|n\times \bar{\alpha}\alpha
}\!\!\!\! &\sim &\!\!\!\!F_{2\kappa +1}^{\sigma _{\alpha }|(\kappa \times 
\bar{\alpha}\alpha )\alpha }(\theta _{1}\ldots \theta _{2\kappa
+1})F_{2(n-\kappa )-1}^{\sigma _{\bar{\alpha}}|[(n-\kappa )\times \bar{\alpha%
}\alpha ]\bar{\alpha}}(\theta _{2\kappa +2}\ldots \theta _{2n}), \\
\bar{\mathcal{T}}_{1,\kappa }^{+}\mathcal{T}_{1,\kappa }^{+}F_{2n+1}^{\sigma
_{\alpha }|(n\times \bar{\alpha}\alpha )\alpha }\!\!\!\! &\sim
&\!\!\!\!F_{2\kappa }^{\mu _{\bar{\alpha}}|\kappa \times \bar{\alpha}\alpha
}(\theta _{1}\ldots \theta _{2\kappa })F_{2(n-\kappa )+1}^{\sigma _{\alpha
}|[(n-\kappa )\times \bar{\alpha}\alpha ]\alpha }(\theta _{2\kappa +1}\ldots
\theta _{2n+1})\,, \\
\bar{\mathcal{T}}_{1,\kappa }^{-}\mathcal{T}_{1,\kappa }^{-}F_{2n+1}^{\sigma
_{\alpha }|(n\times \bar{\alpha}\alpha )\alpha }\!\!\!\! &\sim
&\!\!\!\!F_{2\kappa }^{\mu _{\alpha }|\kappa \times \bar{\alpha}\alpha
}(\theta _{1}\ldots \theta _{2\kappa })F_{2(n-\kappa )+1}^{\sigma _{\alpha
}|[(n-\kappa )\times \bar{\alpha}\alpha ]\alpha }(\theta _{2\kappa +1}\ldots
\theta _{2n+1})\,, \\
\bar{\mathcal{T}}_{1,\kappa }^{+}\mathcal{T}_{1,\kappa
+1}^{+}F_{2n+1}^{\sigma _{\alpha }|(n\times \bar{\alpha}\alpha )\alpha
}\!\!\!\! &\sim &\!\!\!\!F_{2\kappa +1}^{\sigma _{\alpha }|(\kappa \times 
\bar{\alpha}\alpha )\alpha }(\theta _{1}\ldots \theta _{2\kappa
+1})F_{2(n-\kappa )}^{\mu _{\alpha }|(n-\kappa )\times \bar{\alpha}\alpha
}(\theta _{2\kappa +2}\ldots \theta _{2n+1}), \\
\bar{\mathcal{T}}_{1,\kappa }^{-}\mathcal{T}_{1,\kappa
+1}^{-}F_{2n+1}^{\sigma _{\alpha }|(n\times \bar{\alpha}\alpha )\alpha
}\!\!\!\! &\sim &\!\!\!\!F_{2\kappa +1}^{\sigma _{\alpha }|(\kappa \times 
\bar{\alpha}\alpha )\alpha }(\theta _{1}\ldots \theta _{2\kappa
+1})F_{2(n-\kappa )}^{\mu _{\bar{\alpha}}|(n-\kappa )\times \bar{\alpha}%
\alpha }(\theta _{2\kappa +2}\ldots \theta _{2n+1}), \\
\bar{\mathcal{T}}_{1,\kappa }^{+}\mathcal{T}_{1,\kappa }^{+}F_{2n+1}^{\sigma
_{\bar{\alpha}}|(n\times \bar{\alpha}\alpha )\bar{\alpha}}\!\!\!\! &\sim
&\!\!\!\!F_{2\kappa }^{\mu _{\alpha }|\kappa \times \bar{\alpha}\alpha
}(\theta _{1}\ldots \theta _{2\kappa })F_{2(n-\kappa )+1}^{\sigma _{\bar{%
\alpha}}|[(n-\kappa )\times \bar{\alpha}\alpha ]\bar{\alpha}}(\theta
_{2\kappa +1}\ldots \theta _{2n+1})\,, \\
\bar{\mathcal{T}}_{1,\kappa }^{-}\mathcal{T}_{1,\kappa }^{-}F_{2n+1}^{\sigma
_{\bar{\alpha}}|(n\times \bar{\alpha}\alpha )\bar{\alpha}}\!\!\!\! &\sim
&\!\!\!\!F_{2\kappa }^{\mu _{\bar{\alpha}}|\kappa \times \bar{\alpha}\alpha
}(\theta _{1}\ldots \theta _{2\kappa })F_{2(n-\kappa )+1}^{\sigma _{\bar{%
\alpha}}|[(n-\kappa )\times \bar{\alpha}\alpha ]\bar{\alpha}}(\theta
_{2\kappa +1}\ldots \theta _{2n+1})\,, \\
\bar{\mathcal{T}}_{1,\kappa +1}^{+}\mathcal{T}_{1,\kappa
}^{+}F_{2n+1}^{\sigma _{\bar{\alpha}}|(n\times \bar{\alpha}\alpha )\bar{%
\alpha}}\!\!\!\! &\sim &\!\!\!\!F_{2\kappa +1}^{\sigma _{\bar{\alpha}%
}|(\kappa \times \bar{\alpha}\alpha )\bar{\alpha}}(\theta _{1}\ldots \theta
_{2\kappa +1})F_{2(n-\kappa )}^{\mu _{\bar{\alpha}}|(n-\kappa )\times \bar{%
\alpha}\alpha }(\theta _{2\kappa +2}\ldots \theta _{2n+1}), \\
\bar{\mathcal{T}}_{1,\kappa +1}^{-}\mathcal{T}_{1,\kappa
}^{-}F_{2n+1}^{\sigma _{\bar{\alpha}}|(n\times \bar{\alpha}\alpha )\bar{%
\alpha}}\!\!\!\! &\sim &\!\!\!\!F_{2\kappa +1}^{\sigma _{\bar{\alpha}%
}|(\kappa \times \bar{\alpha}\alpha )\bar{\alpha}}(\theta _{1}\ldots \theta
_{2\kappa +1})F_{2(n-\kappa )}^{\mu _{\alpha }|(n-\kappa )\times \bar{\alpha}%
\alpha }(\theta _{2\kappa +2}\ldots \theta _{2n+1}).
\end{eqnarray}
Thus, omitting the shift operators we have formally the following
decomposition of the operators 
\begin{equation}
\Phi _{\alpha }^{\lambda }\longrightarrow \Phi _{\alpha }^{\lambda }\times
\Phi _{\alpha }^{\lambda }\quad \quad \sigma _{\alpha }\longrightarrow
\QATOPD\{ . {\mu _{\alpha }\times \sigma _{\alpha }}{\mu _{\bar{\alpha}%
}\times \sigma _{\alpha }}\quad \quad \mu _{\alpha }\longrightarrow
\QATOPD\{ . {\mu _{\alpha }\times \mu _{\alpha }}{\sigma _{\alpha }\times
\sigma _{\bar{\alpha}}}
\end{equation}
together with the equations for $\alpha \leftrightarrows \bar{\alpha}$. This
means the stated operator content closes consistently under the action of
the cluster decomposition operators.

\section{\protect\bigskip Lie algebraically coupled Federbush models}

The Federbush model as investigated in the previous section only contains
two types of particles. In this section we propose a new Lagrangian, which
admits a much larger particle content. The theories are not yet as complex
as the HSG-models, but they can also be obtained from them in a certain
limit such that they will always constitute a benchmark for these class
theories. Form factors related to these models may be computed similarly as
in the previous section.

Let us consider $\ell \times \tilde{\ell}$-real (Majorana) free Fermions $%
\psi _{a,j}(x)$, now labeled by two quantum \ numbers $1\leq a\leq \ell $, $%
1\leq j\leq \tilde{\ell}$ and described by the Dirac Lagrangian density $%
\mathcal{L}_{\text{FF}}$. We perturb this system with a bilinear term in the
vector currents $J_{a,j}^{\mu }=\bar{\Psi}_{a,j}\gamma ^{\mu }\Psi _{a,j}$ 
\begin{equation}
\mathcal{L}_{\text{CF}}=\sum_{a=1}^{\ell }\sum_{j=1}^{\tilde{\ell}}\bar{\Psi}%
_{a,j}(i\gamma ^{\mu }\partial _{\mu }-m_{a,j})\Psi _{a,j}-\frac{1}{2}\pi
\varepsilon _{\mu \nu }\sum_{a,b=1}^{\ell }\sum_{j,k=1}^{\tilde{\ell}%
}J_{a,j}^{\mu }J_{b,k}^{\nu }\Lambda _{ab}^{jk}\,\,,  \label{CAF}
\end{equation}
and denote the new fields in $\mathcal{L}_{\text{CF}}$ by $\Psi _{a,j}$.
Furthermore, we introduced $\ell ^{2}\times \tilde{\ell}^{2}$ dimensional
coupling constant dependent matrix $\Lambda _{ab}^{jk}$, whose further
properties we leave unspecified at this stage. As in the usual Federbush
model, the effect of the presence of the Levi-Civita pseudotensor $%
\varepsilon $ is that the theory described by $\mathcal{L}_{\text{CF}}$ is
not parity invariant. Thus $\mathcal{L}_{\text{CF}}$ may be viewed as a
system of coupled Federbush models \cite{Feder}.

The formal equations of motion associated to $\mathcal{L}_{\text{CF}}$ are
easily derived as 
\begin{equation}
(i\gamma ^{\mu }\partial _{\mu }-m_{a,j})\Psi _{a,j}=\pi \varepsilon _{\mu
\nu }\gamma ^{\mu }\sum_{b=1}^{\ell }\sum_{k=1}^{\tilde{\ell}}\Lambda
_{ab}^{jk}J_{b,k}^{\mu }\Psi _{a,j}\,.  \label{eqm}
\end{equation}
The solutions to these equations can be constructed in close analogy to the
ones of the Federbush model. The fields 
\begin{equation}
\Psi _{a,j}=\vdots \exp (\sqrt{\pi }i\sum_{b=1}^{\ell }\sum_{k=1}^{\tilde{%
\ell}}\Lambda _{ab}^{jk}\phi _{b,k})\vdots \psi _{a,j}\,=\Phi
_{a,j}^{\lambda }\psi _{a,j}\quad   \label{sol2}
\end{equation}
solve the equations of motion (\ref{eqm}) with the additional assumption
that the bosonic fields $\phi _{a,j}$ constitute potentials for axial vector
currents 
\begin{equation}
\frac{1}{\sqrt{\pi }}\partial _{\mu }\phi _{a,j}=\varepsilon _{\nu \mu
}J_{a,j}^{\nu }=\bar{\psi}_{a,j}\gamma _{\mu }\gamma ^{5}\psi _{a,j},\qquad
\Lambda _{ab}^{jk}\neq 0,\quad \forall \,b,k\,\,.\,  \label{vec2}
\end{equation}
As in the previous section, we used here once again the triple normal
ordering in equation (\ref{sol2}). It needs further computations, similar to
the ones for the Federbush model, to make it rigorous that also in this
context the triple ordering can be associated to a standard Wick normal
ordering. Nonetheless, it appears natural to expect that this can be
generalized analogously and we take this here as an assumption.

Accepting this, we can now compute various equal time exchange relations
with $1\leq a,b\leq \ell $, $1\leq j,k\leq \tilde{\ell}$%
\begin{eqnarray}
\lbrack \phi _{a,j}(x),\phi _{b,k}(y)] &=&[\Phi _{a,j}(x),\Phi _{b,k}(y)]=0
\label{3c2} \\
\lbrack \phi _{a,j}(x),\Phi _{b,k}(y)] &=&\{\psi _{a,j}(x),\psi
_{b,k}(y)\}=0\,\,\,  \label{3d} \\
\lbrack \psi _{a,j}(x),\phi _{b,k}(y)] &=&\sqrt{\pi }\delta _{a,b}\delta
_{j,k}\Theta (x^{1}-y^{1})\psi _{a,j}(x)  \label{333} \\
\psi _{a,j}(x)\Phi _{b,k}^{\lambda }(y) &=&\Phi _{b,k}^{\lambda }(y)\psi
_{a,j}(x)\,e^{-i\pi \Lambda _{ab}^{jk}\Theta (x^{1}-y^{1})} \\
-\psi _{a,j}(x)\Psi _{b,k}(y) &=&\Psi _{b,k}(y)\psi _{a,j}(x)\,e^{-i\pi
\Lambda _{ab}^{jk}\Theta (x^{1}-y^{1})} \\
\Psi _{a,j}(x)\Phi _{b,k}^{\lambda }(y) &=&\Phi _{b,k}^{\lambda }(y)\Psi
_{a,j}(x)\,e^{-i\pi \Lambda _{ab}^{jk}\Theta (x^{1}-y^{1})} \\
-\Psi _{a,j}(x)\Psi _{b,k}(y) &=&\Psi _{b,k}(y)\Psi _{a,j}(x)\,e^{-i\pi
\Lambda _{ab}^{jk}}.  \label{antic2}
\end{eqnarray}

\noindent The equations (\ref{3c2}) and (\ref{3d}) are again clear since $%
\psi _{a,j}$ and $\phi _{a,j}$ are free Fermions and Bosons, respectively.
Equation (\ref{333}) is compatible with (\ref{vec2}) and the remaining
equations are simply consequences of (\ref{3c2})-(\ref{333}). With the help
of these equations we compute directly the scattering matrix. We will be
slightly casual here about complete rigour and do not worry with test
functions and smeared out operators. Noting that $\phi _{a,j}\left|
0\right\rangle =0$, we obtain from 
\begin{equation}
\lim_{t\rightarrow -\infty }\Psi _{a,j}\Psi _{b,k}\left| 0\right\rangle =%
\check{S}_{ab}^{jk}\psi _{a,j}\psi _{b,k}\left| 0\right\rangle \,,\qquad
\lim_{t\rightarrow +\infty }\Psi _{a,j}\Psi _{b,k}\left| 0\right\rangle =%
\hat{S}_{ab}^{jk}\psi _{b,k}\psi _{a,j}\left| 0\right\rangle 
\end{equation}
the S-matrix 
\begin{equation}
S_{ab}^{jk}=(\hat{S}_{ab}^{jk})^{-1}\check{S}_{ab}^{jk}=-e^{i\pi \Lambda
_{ab}^{jk}}\,\,.  \label{S}
\end{equation}
Let us now see whether (\ref{S}) is consistent in the usual sense, i.e. that
it passes all the tests of consistency or if the latter put some constraints
on the possible values for the coupling constant dependent matrix $\Lambda
_{ab}^{jk}$. We demand the usual crossing and unitarity relations (\ref{cu}%
), which means we should have 
\begin{equation}
\Lambda _{ab}^{jk}=-\Lambda _{ba}^{kj}+2\Bbb{Z}\quad \quad \text{and\qquad }%
\Lambda _{ab}^{jk}=\Lambda _{\bar{b}a}^{\bar{k}j}+2\Bbb{Z\,}\,.  \label{con}
\end{equation}
We will now provide some concrete solutions to (\ref{con}) and therefore (%
\ref{S}).

\subsection{HSG-type solutions}

Let us take 
\begin{equation}
\Lambda _{ab}^{jk}=2\lambda _{ab}\varepsilon _{jk}\tilde{I}_{jk}K_{a\bar{b}%
}^{-1}  \label{HSG}
\end{equation}
where $K$ denotes the Cartan matrix of $SU(N)$ and $\tilde{I}$ the incidence
matrix of a simply laced Lie algebra, which we refer to as $\tilde{g}$. The $%
\lambda _{ab}$ are $\ell ^{2}$ coupling constants, which are, however, not
entirely independent of each other. For instance we assume $\lambda
_{ab}=\lambda _{ba}$. Furthermore, we characterise the anti-particle
exclusively by the first quantum number, i.e. $\overline{(a,i)}=(\bar{a},i)$%
, where the particle $\bar{a}$ may be constructed from $a$ by the
automorphism which leaves the associated Dynkin diagram invariant. In the
case of $SU(N)$, we simply have $\bar{a}=\ell +1-a$. It is clear that (\ref
{HSG}) satisfies the first relation in (\ref{con}), whereas the second
relation introduces further constraints on the $\lambda $'s. To be more
concrete we specify now (\ref{S}) \ for some special choices of \ $N\emph{\
\ }$and the Lie algebra$\ \tilde{g}$.

\subsection{The Federbush model}

Considering now the case $SU(3)_{3}$ with $\lambda _{11}=\lambda
_{22}=-2\lambda _{12}=-2\lambda _{21}=\lambda $, we obtain the scattering
matrix of the Federbush model $S^{\text{FB}}$ as defined in (\ref{SFeder}),
where we now used the ordering $\{(1,1),(2,1),(1,2),(2,2)\}$. In comparison
with the previous section, one should notice, that we have now realised this
model in terms real Fermions rather than complex ones.

\subsection{$\tilde{g}_{6}$}

To illustrate the formulae (\ref{con}) and (\ref{S}) a bit more, let us
consider a slighly more complex model, namely $\tilde{g}_{6}$. When
specifying the quantities in (\ref{HSG}) to these algebras, we obtain 
\begin{equation}
S^{ij}=-\left( 
\begin{array}{ccccc}
e^{2\pi i\lambda \varepsilon _{ij}I_{ij}} & e^{2\pi i\lambda ^{\prime
}\varepsilon _{ij}I_{ij}} & 1 & e^{-2\pi i\lambda ^{\prime }\varepsilon
_{ij}I_{ij}} & e^{-2\pi i\lambda \varepsilon _{ij}I_{ij}} \\ 
e^{-2\pi i\lambda ^{\prime }\varepsilon _{ij}I_{ij}} & e^{2\pi i\lambda
^{^{\prime \prime }}\varepsilon _{ij}I_{ij}} & 1 & e^{-2\pi i\lambda
^{^{\prime \prime }}\varepsilon _{ij}I_{ij}} & e^{2\pi i\lambda ^{\prime
}\varepsilon _{ij}I_{ij}} \\ 
1 & 1 & 1 & 1 & 1 \\ 
e^{2\pi i\lambda ^{\prime }\varepsilon _{ij}I_{ij}} & e^{-2\pi i\lambda
^{^{\prime \prime }}\varepsilon _{ij}I_{ij}} & 1 & e^{2\pi i\lambda
^{^{\prime \prime }}\varepsilon _{ij}I_{ij}} & e^{-2\pi i\lambda ^{\prime
}\varepsilon _{ij}I_{ij}} \\ 
e^{-2\pi i\lambda \varepsilon _{ij}I_{ij}} & e^{-2\pi i\lambda ^{\prime
}\varepsilon _{ij}I_{ij}} & 1 & e^{2\pi i\lambda ^{\prime }\varepsilon
_{ij}I_{ij}} & e^{2\pi i\lambda \varepsilon _{ij}I_{ij}}
\end{array}
\right) \,,
\end{equation}
where the rows and columns are ordered as $1,2,\ldots 5$. In this case, we
have three independent coupling constants $\lambda ,\lambda ^{\prime }$ and $%
\lambda ^{\prime \prime }$.

\subsection{The HSG-limit}

From the specific example in (\ref{SFeder}), we expect that the HSG-models
are in general closely related to (\ref{S}) with (\ref{HSG}). Indeed, taking 
$\lambda _{ab}=1$ for $1\leq a,b\leq \ell $, we obtain 
\begin{equation}
S_{ab}^{jk}=-e^{2\pi i\varepsilon _{jk}I_{jk}K_{a\bar{b}}^{-1}}\,\,,
\end{equation}
which clearly satisfies the first relation in (\ref{con}), whereas for the
second relation, we simply have to recall the well-known fact that $%
(K_{SU(N)}^{-1})_{ab}=\min (a,b)-ab/N$. Comparing with the expression 
\begin{equation}
\lim_{\theta \rightarrow \infty }[S_{ab}^{jk\,\text{HSG}}(\theta
)S_{ab}^{jk\,\text{HSG}}(-\theta )]=e^{2\pi i\,(K_{\bar{a}b}^{\text{SU(N)}%
})^{-1}I}\,\,.
\end{equation}
we note that these solutions coincide. This means when one eventually solves
the HSG-models, one can always take the limit to the corresponding quantities 
of $\mathcal{L}_{\text{CF}}$ for consistency checks. 

\section{The ultraviolet limit}

When having found a solution to the form factor consistency equations, with
the factor of local commutativity and the scattering matrix as the only
input, one normally does not know which operator this particular solution
corresponds to. Of course in the present situation we are in a better
position, since we are already working with an explicit representation for
the operators. Nonetheless, in section 4.2.2. we saw that even this can
still lead to wrong assignments and it is desirable to have more
information. By calling the operators $\Phi _{\alpha }^{\lambda },\mu ,$ and 
$\Sigma _{\alpha }^{\lambda },\sigma $, disorder and order operators,
respectively, we have already borrowed the terminology from the underlying
conformal field theory. In order to make this correspondence more manifest
one may carry out explicitly the ultraviolet limit. The ultraviolet Virasoro
central charge of the theory itself can be computed from the knowledge of
the form factors of the trace of the energy-momentum tensor \cite{ZamC} by
means of the expansion 
\begin{equation}
c_{\text{uv}}=\sum_{n=1}^{\infty }\sum_{\mu _{1}\ldots \mu _{n}}\frac{9}{%
n!(2\pi )^{n}}\int\limits_{-\infty }^{\infty }\ldots \int\limits_{-\infty
}^{\infty }\frac{d\theta _{1}\ldots d\theta _{n}}{\left(
\sum_{i=1}^{n}m_{\mu _{i}}\cosh \theta _{i}\right) ^{4}}\left|
F_{n}^{T_{\;\;\mu }^{\mu }|\mu _{1}\ldots \mu _{n}}(\theta _{1},\ldots
,\theta _{n})\right| ^{2}\,.  \label{ccc}
\end{equation}
In a similar way one may compute the scaling dimension of the operator $%
\mathcal{O}$ from the knowledge of its n-particle form factors \cite{DSC} 
\begin{eqnarray}
\Delta _{\text{uv}}^{\!\!\mathcal{O}} &=&-\frac{1}{2\left\langle \mathcal{O}%
\right\rangle }\sum_{n=1}^{\infty }\sum_{\mu _{1}\ldots \mu
_{n}}\int\limits_{-\infty }^{\infty }\ldots \int\limits_{-\infty }^{\infty }%
\frac{d\theta _{1}\ldots d\theta _{n}}{n!(2\pi )^{n}\left(
\sum_{i=1}^{n}m_{\mu _{i}}\cosh \theta _{i}\right) ^{2}}  \nonumber \\
&&\times F_{n}^{T_{\;\;\mu }^{\mu }|\mu _{1}\ldots \mu _{n}}(\theta
_{1},\ldots ,\theta _{n})\,\left( F_{n}^{\mathcal{O}|\mu _{1}\ldots \mu
_{n}}(\theta _{1},\ldots ,\theta _{n})\,\right) ^{\ast }\,\,\,.
\label{dcorr}
\end{eqnarray}
In general the expressions (\ref{ccc}) and (\ref{dcorr}) yield the
difference between the corresponding infrared and ultraviolet values, but we
assumed here already that the theory is purely massive such that the
infrared contribution vanishes. Let us now evaluate these formulae.

\subsection{The complex Fermion}

Since for the free Fermion one only has to sum up to the two particle
contribution, the infinite sum (\ref{ccc}) and (\ref{dcorr}) terminate and
can be evaluated even analytically. For the case $N=2$ we obtain 
\begin{equation}
c_{\text{uv}}=2\qquad \text{and\qquad }\Delta _{\text{uv}}^{\!\!\mu _{\alpha
}}=\Delta _{\text{uv}}^{\!\!\mu _{\bar{\alpha}}}=\frac{1}{16}\,.  \label{16}
\end{equation}
The scaling dimensions of $\sigma _{\alpha }$ and $\sigma _{\bar{\alpha}}$,
which are expected to coincide with (\ref{16}), can not be computed from (%
\ref{dcorr}), since it involves an odd number of particles.

\subsection{The Federbush model}

We may proceed similarly for the Federbush model. In the ultraviolet limit
it obviously corresponds to two complex free Fermions and we there expect to
obtain 
\begin{equation}
c_{\text{uv}}=2\,\,.
\end{equation}
Indeed using (\ref{EMTF}), the computation is identical to the one carried
out in the previous section. Note, that this value of $2$ coincides with the
ultraviolet central charge of the $SU(3)_{3}$-HSG model. This is however
also not entirely surprising by recalling the identification (\ref{SFeder}).
The corresponding thermodynamic Bethe ansatz equations will be identical to
for free Fermions. More striking is the result of the evaluation of (\ref
{dcorr}), which yields with (\ref{EMTF}) and (\ref{phi}) 
\begin{equation}
\Delta _{\text{uv}}^{\!\!\Phi _{\alpha }^{\lambda }}=\Delta _{\text{uv}%
}^{\!\!\Phi _{\alpha }^{\lambda }}=\frac{\lambda ^{2}}{4}\,.  \label{lam}
\end{equation}
Note, that $\Delta _{\text{uv}}^{\!\!\Phi _{\alpha }^{1/2}}=\Delta _{\text{uv%
}}^{\!\!\Phi _{\alpha }^{1/2}}=1/16$, which is once again the limit to the
complex free Fermion. Yet more support for the relation between the $%
SU(3)_{3}$-HSG model and the Federbush model comes from the analysis of $%
\lambda =2/3$, which corresponds to the $SU(3)_{3}$-HSG value $\tau =1/3$
(see (\ref{62})). In that case we obtain from (\ref{lam}) $\Delta _{\text{uv}%
}^{\!\!\Phi _{\alpha }^{2/3}}=\Delta _{\text{uv}}^{\!\!\Phi _{\alpha
}^{2/3}}=1/9$. We now compare with the general formula for the scaling
dimensions of the $SU(3)_{3}$-HSG model 
\begin{equation}
\Delta (\Lambda ,w)=\frac{(\Lambda \cdot (\Lambda +2\rho ))}{12}-\frac{%
(w\cdot w)}{6}\,\,,  \label{wei}
\end{equation}
where $\Lambda $ is a highest weight vector of level smaller or equal $3$, $%
w $ the corresponding lower weights and $\rho $ the Weyl vector. We are
specially interested in the field corresponding to $\Delta (\lambda
_{1},\lambda _{1})$ with $\lambda _{1}$ being a fundamental weight, since
this field was previously observed \cite{CF1} to correspond to the disorder
operator. Indeed, we find that 
\begin{equation}
\Delta (\lambda _{1},\lambda _{1})=\Delta _{\text{uv}}^{\!\!\Phi _{\alpha
}^{2/3}}=\Delta _{\text{uv}}^{\!\!\Phi _{\alpha }^{2/3}}\,.
\end{equation}
Thus precisely at the value of the coupling constant of the Federbush model
at which the $SU(3)_{3}$-HSG S-matrix reduces in the limit (\ref{SFeder}) to
the $S^{\text{FB}}$, the operator content of the two models overlaps.

\section{Conclusions}

We computed explicitly the form factors of the complex free Fermion and the
Federbush model related to various operators. On one hand we carried out
this task by representing explicitly the field content as well as the
particle creation operator in terms of fermionic Fock operators and computed
thereafter directly the corresponding matrix elements. On the other hand we
verified that these expressions satisfy the form factor consistency
equations only when the operators under consideration are mutually local.
This can already be seen for the free Fermion, for which we could have also
computed the matrix element of the field $\Phi _{\alpha }^{\lambda }(x)$. In
that context one observes that only for $\lambda =1/2$ this function solves
the consistency equations (\ref{W1})-(\ref{kin}). We observed a similar
phenomenon in the Federbush model. Whereas the matrix elements of the field $%
\Sigma _{\alpha }^{\lambda }(x)$ can be computed in a closed form for
generic values of $\lambda $, they become only meaningful form factors for $%
\lambda =1/2$, that is when the field becomes local. It turned out to be
crucial that the consistency equations contain the factor of local
commutativity as defined in (\ref{local}). It is important to note that this
factor is related to the equal time exchange relation between the operator
and the field associated to the particle next to it in the multi-particle
state.

Our solutions turned out to decompose consistently under the momentum space
cluster property.

Further support for the identification of the solutions was given by an
analysis of the ultraviolet limit.

We demonstrated how the scattering matrix of the Federbush model can be
obtained as a limit of the $SU(3)_{3}$-HSG scattering matrix. This
``correspondence'' also holds for the central charge, which equals $2$ in
both cases, and the scaling dimension of the disorder operator at a certain
value of the coupling constant.\ We proposed a Lie algebraic generalization
of the Federbush models, which on the other hand can be obtained in a
certain limit of the homogeneous sine-Gordon models.

We expect that the construction of form factors by means of free fermionic
Fock fields can be extended to other models.\medskip

\noindent \textbf{Acknowledgments: } We are grateful to the Deutsche
Forschungsgemeinschaft (Sfb288), for financial support and to
J.L.~Miramontes, D.I.~Olive, S.~Pakuliak and B. Schroer for useful comments.
Special thanks to T.D.~Crawford and E.A.~Crawford for sharing their Wick
contraction style files with us.

\end{document}